\newcommand\yash[1]{\textbf{\textcolor{violet}{Yash: #1}} }

\documentclass[letterpaper,twocolumn,10pt]{article}
\usepackage{multirow}
\usepackage{usenix,soul}

\usepackage{tikz}
\usepackage{amsmath}
\usepackage{float}
\usepackage{stfloats}

\usepackage{filecontents}

\microtypecontext{spacing=nonfrench}

\usepackage{tcolorbox}
\usepackage{colortbl}
\usepackage{xcolor}
\usepackage{array}
\usepackage{booktabs}
\usepackage{makecell}
\usepackage{adjustbox}
\usepackage{amssymb}
\usepackage{tabularx}

\usepackage{multirow}
\usepackage{graphicx}
\usepackage{fontawesome5} 
\usepackage{pifont} 
\usepackage{bbding}
\usepackage{caption}
\usetikzlibrary{shapes, calc} 
\newcommand{\circleFull}{\tikz\draw[fill=black] (0,0) circle (0.1cm);} 
\newcommand{\circleNone}{\tikz\draw (0,0) circle (0.1cm);}            
\newcommand{\circleLeft}{\tikz\draw[fill=black] (0,0) circle (0.1cm) (0,0) -- ++(90:0.1cm) arc[start angle=90, end angle=-90, radius=0.1cm] -- cycle;}
\newcommand{\circleRight}{\tikz\draw[fill=black] (0,0) circle (0.1cm) (0,0) -- ++(270:0.1cm) arc[start angle=270, end angle=90, radius=0.1cm] -- cycle;} 
\newcommand{\HalfCheckmark}{{\ding{52}\textsuperscript{\kern-0.5em\small\ding{56}}}}
\newcommand{\RestrictionSymbol}{
    \tikz[scale=0.28]{
        \draw[line width=0.6mm] (0,0) circle(0.5); 
        \draw[line width=0.6mm] (-0.35,-0.35) -- (0.35,0.35); 
    }
}

\newcommand{\yellowbox}[1]{\adjustbox{margin=2px,bgcolor={HTML}{fcd14f},padding=1px}{#1}}
\newcommand{\redbox}[1]{\adjustbox{margin=2px,bgcolor={HTML}{ff9999},padding=1px}{#1}}
\newcommand{\graybox}[1]{\adjustbox{margin=2px,bgcolor={HTML}{c0c0c0},padding=1px}{#1}}

\newcommand{\Exfilredbox}[1]{\colorbox{CBRed!75}{\strut#1}}
\newcommand{\Exfilyellowbox}[1]{\colorbox{CBYellow!75}{\strut#1}}
\newcommand{\Exfilorangebox}[1]{\colorbox{CBOrange!75}{\strut#1}}
\newcommand{\Exfilgreenbox}[1]{\colorbox{CBGreen!75}{\strut#1}}

\definecolor{lightred}{rgb}{1.0, 0.8, 0.8}
\definecolor{lightyellow}{rgb}{1.0, 1.0, 0.8}
\definecolor{lightblue}{rgb}{0.8, 0.9, 1.0}
\definecolor{lightgrey}{rgb}{0.9, 0.9, 0.9}

\begin{document}


\title{\Large \bf Big Help or Big Brother?\\
Auditing Tracking, Profiling, and Personalization in Generative AI Assistants}

\author{
{\rm Yash Vekaria}\\ UC Davis\\
\and
{\rm Aurelio Loris Canino}\\ UNIRC\\
\and
{\rm Jonathan Levitsky}\\ UC Davis\\
\and
{\rm Alex Ciechonski}\\ UCL\\
\and
{\rm Patricia Callejo}\\ UC3M\\
\and
{\rm Anna Maria Mandalari}\\ UCL\\
\and
{\rm Zubair Shafiq}\\ UC Davis\\
} 

\maketitle

\begin{abstract}
Browser assistants have started to integrate powerful capabilities of GenAI in web browsers to offer functionalities such as question answering, content summarization, and agentic web navigation. 
These assistants, available today as browser extensions, raise significant privacy concerns because they can track detailed browsing activity (e.g., searches, clicks) and autonomously perform tasks such as form filling.
In this paper, we analyze the design and behavior of GenAI browser extensions, focusing on how they collect, process, and share user data, and whether they profile users based on explicit or inferred demographic attributes and interests.
We develop a novel prompting framework and perform network traffic analysis to audit the nine GenAI browser assistants for tracking, profiling, and personalization.

We find that GenAI browser assistants typically rely on server-side APIs rather than local models, and can be invoked automatically without explicit user interaction.
GenAI browser assistants often collect and share full webpage content, including the HTML DOM and user form inputs in some cases, with their first-party servers.
Some also share identifiers and user prompts with third-party trackers such as Google Analytics.
This data collection and sharing happens even on pages containing sensitive information, such as health records or personal information such as names or social security numbers entered in a web form.
Moreover, several GenAI browser assistants infer attributes (e.g., age, gender, income, interests) and use them to personalize responses across browsing contexts.
Our findings show that GenAI browser assistants collect and share personal and sensitive information for profiling and personalization, highlighting the need for safeguards as they increasingly mediate web browsing. 
\end{abstract}

\section{Introduction}
\label{sec:introduction}

With recent advances in natural language processing, Large Language Models (LLMs) trained on massive datasets and built with billions of parameters have significantly improved their ability to understand context and generate content across modalities, including text, images, and videos~\cite{minaee2024large}.
As the foundational technology behind many Generative Artificial Intelligence (GenAI) systems, LLMs now power a wide range of applications such as chatbots and workflow automations. 
Popular search engines have also begun integrating LLMs. 
For instance, Google Search now displays Gemini-generated overviews in response to user queries \cite{Reid_2024}, while Microsoft Bing's Copilot uses models from OpenAI \cite{Mehdi_2023}. 
However, these search engines are limited to analyzing activity on their own platforms, restricting their visibility of a user's browsing activity across the web. 
To address this limitation, GenAI browser assistants have emerged to use LLMs to enhance a user's overall browsing experience across the web.

GenAI browser assistants are implemented as browser extensions, giving them access to nearly everything a user does in their browser.
By operating as a browser extension, these assistants can track a user's browsing activity across websites such as web pages visited, buttons clicked, search terms and other personal information typed into form fields.
This browsing activity provides context that assistants can use to build detailed user profiles.
While this enables personalized responses by these assistants, it also raises serious privacy concerns.
For example, browsing data and user profiles can be collected, shared, and repurposed for other usecases such as online behavioral advertising. 
Despite high operational costs~\cite{Landymore_2023}, most LLM platforms do not yet rely on ads for monetization. 
However, this shift is imminent \cite{Perplexity_Team_2024,arstechnica2025,ft2025}.

%
%

It is important to understand the privacy risks posed by tracking, profiling, and personalization by GenAI-based browser assistants. 
Millions of users already use these assistants, yet little is known about how they are designed, what information they collect and share, and how they process it.
To this end, we propose a novel framework to systematically audit GenAI browser assistants, guided by the following research questions:

\noindent {\textbf{RQ1.\textit{ How is the architecture of GenAI browser assistants designed?}}} 
We analyze network traffic of these browser assistants to examine key design choices: (1) backend model capabilities, (2) response architecture, (3) context isolation across interactions, and (4) response variability.

\noindent {\textbf{RQ2. \textit{Do GenAI browser assistants collect and share user information?}}} 
We analyze network traffic of these browser assistants to examine how user data is handled, focusing on: (1) implicit and explicit collection and sharing, (2) of data with first-party and third-party servers, and (3) differences across public and private online spaces.

\noindent {\textbf{RQ3. \textit{Do GenAI browser assistants profile users based on their browsing behavior to personalize responses}}}? 
We develop a prompting framework to evaluate profiling and personalization, focusing on: (1) whether assistants infer user attributes such as location, age, gender, income, and interests from browsing activity, and (2) whether they use this information to personalize responses within and across browsing contexts.

We summarize our key contributions and findings below:

\noindent $\bullet$ \textbf{Architecture.} We analyze nine popular GenAI browser assistants and find that all but one rely on server-side response generation; only a single assistant performs response generation on the client side.
    Some assistants are automatically invoked when a user enters a query on a search engine.
    Furthermore, seven assistants isolate context across browsing sessions and tabs, while the remaining two do not do so.

\noindent $\bullet$ \textbf{Tracking.}
    Using assistants in private online spaces resulted in data collection ranging from partial webpage content to full DOM snapshots.
    Merlin was observed to extract form inputs, leading to the collection and sharing of sensitive information, including medical records from a university health portal (\url{hem.ucdavis.edu}), academic records from Canvas (from \url{canvas.edu}), and a Social Security number entered on an IRS website \url{irs.gov}). 
    Sider and Merlin shared chat and user identifiers with \url{google-analytics.com} while TinaMind shared them with \url{analytics.google.com}. 
    Merlin also shared raw user queries with Google Analytics, which could potentially be used for tracking and personalization across Google platforms. 
    Chat histories were shared with first-party servers by four assistants. 
    Moreover, Harpa and Copilot stored the complete chat history in the extension's background service worker using IndexedDB, indicating that these histories persist across browsing sessions.
    
\noindent $\bullet$ \textbf{Profiling and Personalization.} Two extensions (Monica, and Sider) demonstrated profiling based on all five user attributes (location, age, gender, income, and interests) and responded to both in-context and out-of-context personalization prompts. 
In contrast, Perplexity and TinaMind showed no strong evidence of profiling or personalization, while Harpa exhibited only in-context personalization behavior.


\section{Background}
\label{sec:background}

We provide background of GenAI in Section~\ref{sec:background:genAI-systems} and its application in browser assistants in Section~\ref{sec:background:genAI-browser-assistants}.

\subsection{Generative AI}
\label{sec:background:genAI-systems}


\noindent \textbf{{How do GenAI systems work?}}
Generative AI (GenAI) systems typically rely on large language models (LLMs), which operate by iteratively predicting the next token in a sequence based on the context of preceding tokens~\cite{Zewe2023}.
These models are built on transformer architectures, where a self-attention mechanism enables the model to assign varying importance to different parts of the input sequence~\cite{vaswani2017attention}.
When a user inputs a query, the model encodes the input into an embedding space and generates a probability distribution over possible next tokens.
It then selects a token based on this distribution and repeats the process until the output is complete.
This approach enables GenAI systems to provide coherent and context-aware responses to user input.

\noindent \textbf{{Limitations and challenges of GenAI systems}.} 
GenAI systems face several challenges that affect their performance, usability, and broader applicability.
Understanding these limitations is essential for evaluating how these systems function in practice.
The first fundamental issue is their \textit{probabilistic} nature. 
As described earlier, GenAI systems generate the next token based on a probability distribution, which can lead to varying outputs even for the same input. 
This variability is influenced by the \textit{temperature} parameter, which controls the randomness of the output. 
Higher temperature values encourage the selection of less probable tokens, often producing more creative but potentially less coherent responses, while lower values lead to more deterministic and focused responses \cite{minaee2024large}.
The second challenge is the resource intensive nature of training these models \cite{Buchholz_2024}. 
This means that these models cannot incorporate up-to-date information that was not part of the training process without access to live search capabilities. 
Finally, while general-purpose models are versatile, adapting them to domain-specific use cases, such as personalized assistants is non-trivial. 
These typically have access to only the user-submitted queries within the platform, which limits their context to what is explicitly included in the user prompt and makes personalization challenging.

\vspace{-3mm}
\subsection{Generative AI Browser Assistants}
\label{sec:background:genAI-browser-assistants}

\noindent \textbf{{How do GenAI browser assistants function?}} 
Gen AI browser assistants aim to leverage capabilities of GenAI models to personalize user experience on the web. 
Implemented as browser extensions, these assistants function as wrappers around GenAI models such as OpenAI's ChatGPT~\cite{chatgpt2025}, Google's Gemini~\cite{gemini2025}, Meta's Llama~\cite{llama2025}, etc. 
They monitor user activity on the web and use this context to query backend GenAI APIs and return personalized responses.
Unlike foreground web applications, these assistants operate through background service workers that run continuously while the extension is active.
They can inject \textit{content scripts} into every page that a user visits.
These content scripts may load additional libraries (e.g., jQuery) to facilitate actions like DOM manipulation, and can communicate with the background script to log and share user activity.
This browser extension based architecture effectively grants the assistants broad access to everything a user does in the browser.
By monitoring a user's browsing activity across websites, browser assistants can offer in-the-moment suggestions tailored to specific browsing contexts.
In summary, these assistants address many of the limitations of general-purpose GenAI systems and offer a more personalized, context-aware experience.

\noindent \textbf{{What functionalities do GenAI browser assistants offer?}} 
GenAI browser assistants offer a range of features that we group into three broad categories: search-based assistants, content-based assistants, and automation assistants.
Search-based assistants offer features such as natural language responses to search queries, support web page summarization, interactive Q\&A with web pages, and text highlighting. 
Content-based assistants assist with content-specific tasks such as writing tasks, social media post generation, meeting transcript summarization, YouTube video explanation, SEO generation, etc.
Automation assistants focus on streamlining workflows by supporting tasks such as auto-filling forms, coding and debugging agents, task scheduling, automated data extraction and scraping, voice-enabled assistance, etc.

\vspace{-3mm}
\subsection{Traditional vs. GenAI Browser Extensions}
\label{sec:background:traditional-genai-extensions}

While browser extensions have long been supported in browsers, the integration of GenAI in browser extensions introduces new risks through expanded data collection and inference capabilities.
To understand these risks, we highlight three key differences between traditional and GenAI-powered browser extensions:

\noindent {\textbf{Third-party model integration.}} 
GenAI extensions rely on third-party GenAI models to analyze a user's browsing activity.
This allows them to build rich user profiles by combining observed browsing behavior with inferred characteristics such as interests and intent. 
Distinct from traditional extensions, GenAI extensions can personalize responses based not only on what users explicitly do, but also on what the LLM infers about them through reasoning.

\noindent {\textbf{Extensive data collection.}} 
Traditional extensions typically collect limited, task-specific data such as URLs or selected DOM elements. 
For instance, a price comparison extension may only extract product names and prices from searches on a specific site like Amazon.
On the contrary, GenAI extensions often collect and transmit full-page content or the entire DOM. 

\noindent {\textbf{Memory.}} 
Many GenAI browser assistants include a memory component that persists across navigations, sessions, or tabs.
This enables longitudinal tracking and cross-context profiling, which is not common in traditional extensions.

\vspace{-3mm}
\section{Related Work}
\label{sec:related-works}
\vspace{-2mm}

In this section, we review related literature along two dimensions: privacy issues in browser extensions and privacy issues in GenAI. 
We conclude by discussing how our work compares to and builds upon prior research in this space.

\noindent {\textbf{Privacy issues in browser extensions}.} 
While browser extensions add to functionality, they pose significant privacy risks due to their access to user data such as browsing history, local storage, payment information, and more~\cite{rana2014security}. 
These risks stem from the potential exploitation of extensions by malicious actors seeking unauthorized access to user data~\cite{fass2021doublex, 8947709}. 
Prior research has examined various aspects of these privacy risks such as detecting spying extensions~\cite{8406590}, identifying inconsistencies in privacy practices~\cite{bui2023detection}, and finding data leaks~\cite{xie2020jtaint}. 
There has also been some work from the security perspective, focusing on detecting vulnerable and malicious extensions \cite{shahriar2014effective}, enabling cloud-based security analysis~\cite{das2016cloud}, and security analysis of sensitive data access~\cite{nayak2024experimental, bandhakavi2011vetting}.
Chen et al.~\cite{chen2018mystique} found that 2.13\% of Chrome browser extensions, representing over 60 million users, leak sensitive information such as browsing history, open tabs, passwords, and location data. 
Xie et al.~\cite{xie2024arcanum} further further identified hundreds of extensions that automatically extract user content from web pages~\cite{martin2001privacy}. 
Well-known examples include popular extensions like PayPal’s Honey, Capital One Shopping, Hola VPN, and Avira Safe Shopping, which exfiltrate URLs and other user data. 
Alarmingly, around 40\% of Chrome extensions, including the widely used ones, exhibit security vulnerabilities that could compromise user data~\cite{8947709}.
Moreover, Ling et al.~\cite{ling2022they} found that 92\% of browser extensions collect data in ways that contradict their stated privacy policies. 
Many extensions were found to collect excessive data beyond what was necessary for their intended functionality. 
For instance, while the Chrome extension \textit{InserLearning} claimed to only collect a user's name, Google account email, and profile image, but was also found to collect full webpage content. 
Additionally, Carnus et al.~\cite{carnus2020} demonstrated that some extensions leak user data, such as email addresses, names, and phone numbers, to their developers and third parties.

\noindent \textbf{{Privacy issues in GenAI models}.} 
Past work in this space has mostly focused on investigating privacy issues resulting from the training phase of GenAI models.
Researchers have demonstrated that LLMs can overfit to sensitive information within training datasets, potentially resulting in the unintended exposure of private data during interactions through APIs, browser extensions, or other interfaces \cite{gupta2023chatgpt, li2023privacy, li2024llm, kibriya2024privacy, wang2023decodingtrust, peris2023privacy}. 
Such risks are amplified by the susceptibility of LLMs to various attack vectors, including model inversion attacks (reconstruction of sensitive data from model outputs), adversarial attacks (e.g., prompt injection), and data poisoning attacks (introduction of malicious data into training sets) \cite{yu2023gptfuzzer, bowen2024data, zhou2024model}.
To mitigate these risks, prior research has investigated privacy-preserving methods such as differential privacy, federated learning, and secure multi-party computation aimed at reducing the exposure of sensitive user information \cite{yan2024protecting}. 
There remain critical gaps to prevent GenAI models from learning sensitive attributes during the inference phase. 
Staab et al. show that even limited data from users’ online activities on platforms like Reddit or Twitter enables an LLM to infer a user's personal attributes with 85\% accuracy \cite{staab2023beyond}.

\noindent \textbf{{Comparison with prior work}.} 
Past research has primarily focused on privacy-invasive data collection by browser extensions and the associated security vulnerabilities that expose user data to malicious actors. 
However, GenAI browser extensions integrated with LLMs introduce new risks, as discussed in Section~\ref{sec:background:traditional-genai-extensions}.
These risks include potential exposure of browsing activity, form inputs and other interaction to LLMs, which are highly effective at inferring sensitive user attributes from seemingly benign interactions \cite{staab2023beyond}. 
Our work builds on the findings of Staab et al.~\cite{staab2023beyond} by examining the compounded privacy risks of combining LLMs with browser extension capabilities, an area that has been understudied. 
To the best of our knowledge, ours is the first systematic audit focused specifically on GenAI browser assistants.
While prior work has analyze collection and sharing of user data, they have not investigated how that data is subsequently used. 
GenAI browser assistants are a new target for privacy analysis, allowing us to investigate not just data collection and sharing but also how it may be used for profiling and personalization. 
We investigate whether continuous tracking of user activities within the browser allows assistants to associate memory with a user's personal and behavioral traits, using updated context to personalize responses over time.
Being a fairly new and developing space, it is both important and timely to investigate the unique privacy risks they pose. 
To that end, we design a novel experimental framework to audit tracking, profiling, and personalization across three major functionalities (search, browsing, and summarization) offered by nine of the most widely used GenAI browser assistants.

\vspace{-3mm}
\section{Threat Model}
\vspace{-2mm}

Our threat model considers online users who install a GenAI browser assistant to personalize their browsing experience.
In this threat model, the browser extension of the assistant is treated as the adversary while the user is the victim. 
The adversary operates as a browser extension, which provides it access to a user's complete browsing activity. 
The primary goal of the adversary is to provide personalized responses to user queries. 
To achieve this, it is assumed to engage in user tracking by monitoring browsing activities and extracting contextual information.
This user tracking is considered the secondary (or implicit) goal of the adversary. 
Adversary may share the collected information from the victim's browser either to its own servers or to third-party servers.

\begin{figure}[!t]
    \centering
\includegraphics[width=\linewidth]{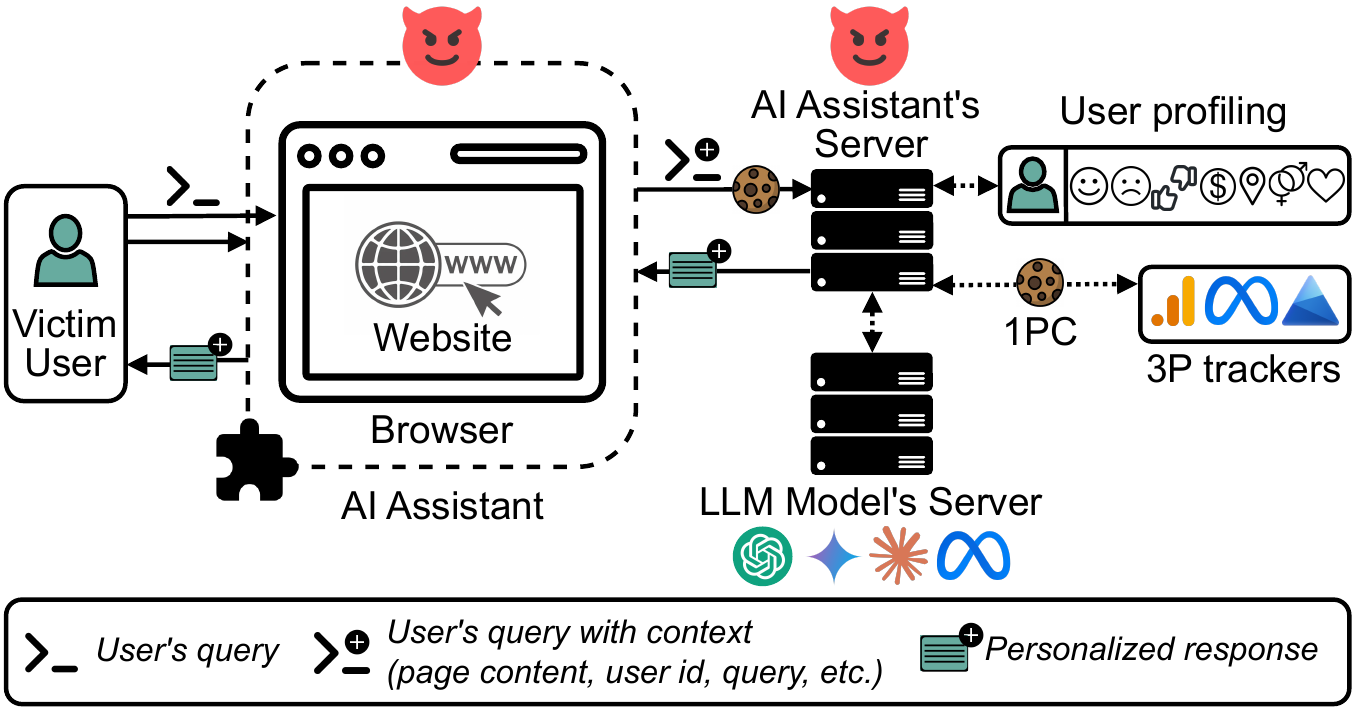}
    \vspace{-5mm}
    \caption{Threat model of GenAI browser assistants}
    \label{fig:threat-model}
   \vspace{-5mm}
\end{figure}

Figure~\ref{fig:threat-model} illustrates our threat model.
The victim can interact with the GenAI browser assistants in different ways such as initiating a Google search, summarizing a web page, chatting about the web page's content, etc.
Any such activity that involves a user's interaction with the GenAI assistant is assumed to be a user's query. 
Upon receiving a user query, the assistant sends it along with metadata to its own backend server.
The metadata may include user-, device-, webpage-, or browsing-specific information that serves as a \textit{context} for generating a personalized response.
The assistant’s server may generate the response internally or share the query and context to an external LLM API. 
The personalized response is sent back to the victim's browser.

The assistant may also set and share cookies from the victim's browser to its own server. 
These cookies could further be shared with third-party advertising or analytics platforms to support analytics or ad targeting, such as building custom audiences for personalized advertising.
To fulfill its primary goal, the adversary is assumed to continuously build and refine a profile of the user based on accumulated context over time.
We consider the LLM provider and third-party trackers as beneficiaries of the data collected and shared by the assistant.
Although they may gain this information, we do not assume them to be directly involved in data collection.


\begin{table*}[ht!]
\centering
\small
\renewcommand{\arraystretch}{1.5} 
\setlength{\tabcolsep}{5pt} 

\caption{Overview of studied AI browser assistants sorted by their popularity.} \vspace{-3mm} \textit{Legend}: 
\protect\redbox{Personal Data}: Personally Identifiable Information (\textbf{PII}), Personal Communications (\textbf{PC}), Financial Information (\textbf{FI}). \\
\protect\yellowbox{Web Data}: User Activity (\textbf{UA}), Web History (\textbf{WH}), Website Content (\textbf{WC}), Location (\textbf{LOC}). 
\protect\graybox{No Data}: \textbf{No Data} collected.\\

\label{tab:extension-details}
\begin{tabular}{@{}p{3.6cm}p{1cm}p{1.9cm}p{2.0cm}p{1.4cm}p{1.5cm}p{2.88cm}p{1.1cm}@{}}

\toprule
\textbf{Extension$\newline$Name} & \textbf{Install$\newline$Counts} & \textbf{Supported Model(s)} & \textbf{Default$\newline$Model} & \textbf{Invocation Mode} & \textbf{Response Mode}  & \textbf{Data$\newline$Disclosures} & \textbf{SDK Version} \\ 
\midrule

\textbf{Sider}: ChatGPT Sidebar 
& 5M 
& \includegraphics[width=0.035\textwidth, height=0.02\textwidth]{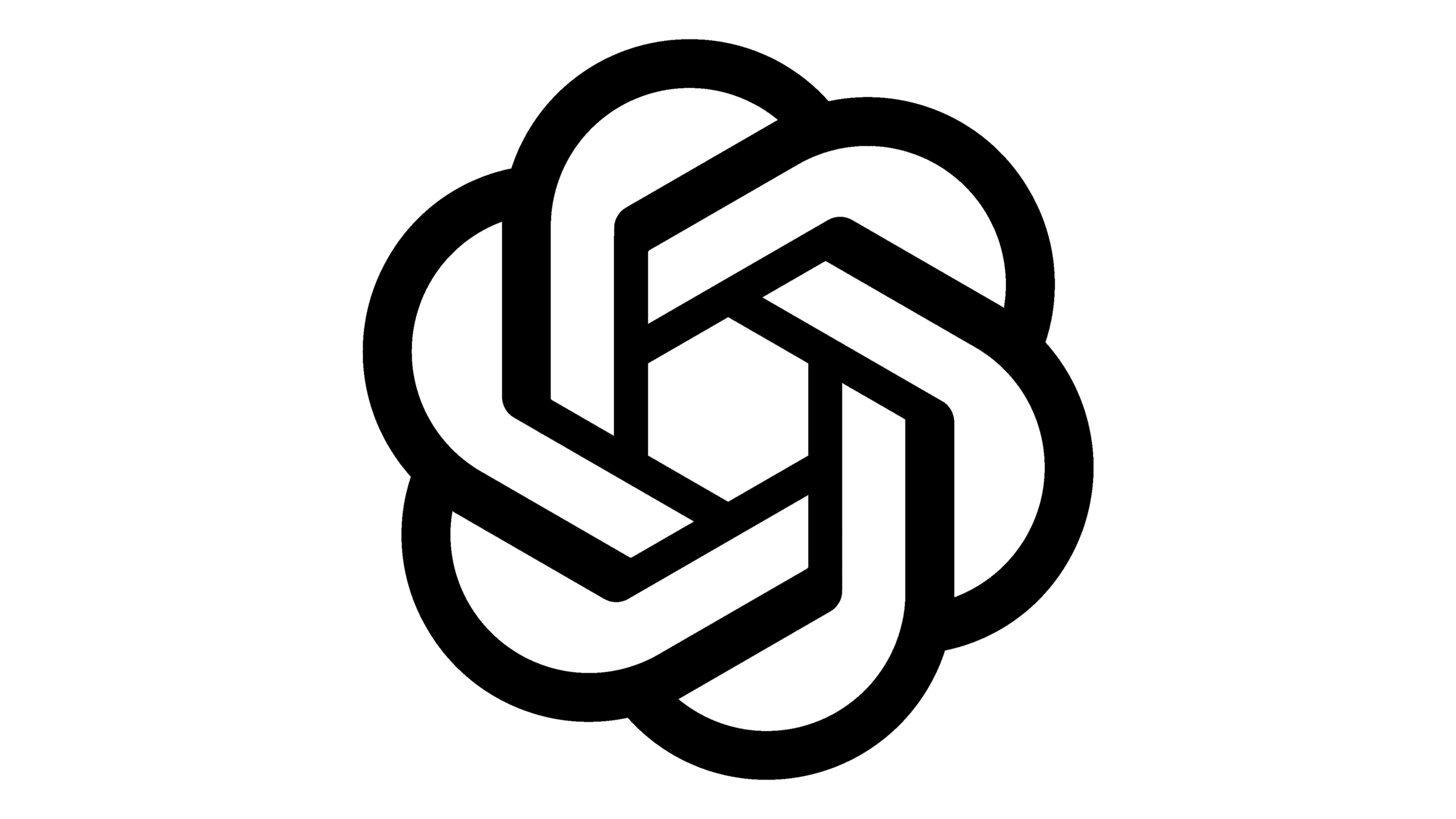}\includegraphics[width=0.017\textwidth, height=0.02\textwidth]{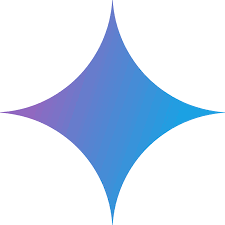} \includegraphics[width=0.02\textwidth, height=0.02\textwidth]{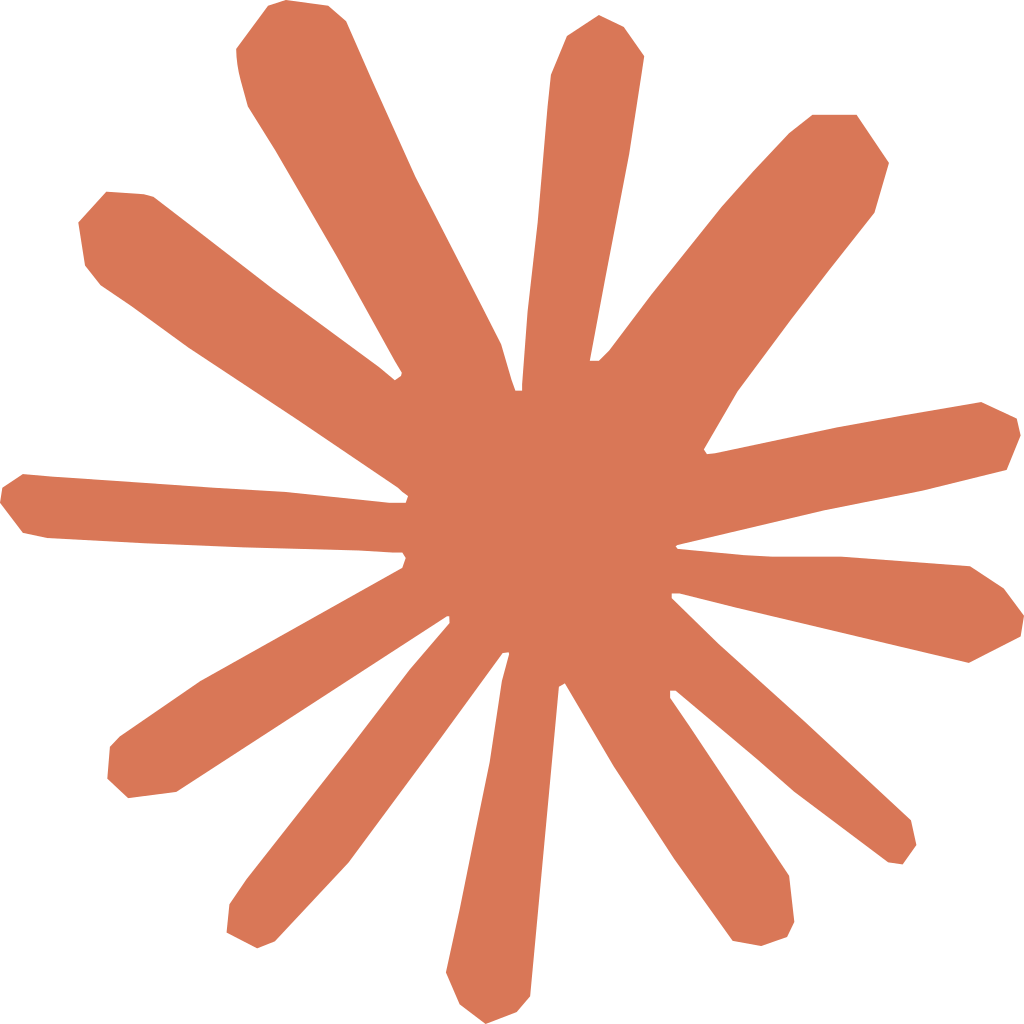} \includegraphics[width=0.022\textwidth, height=0.017\textwidth]{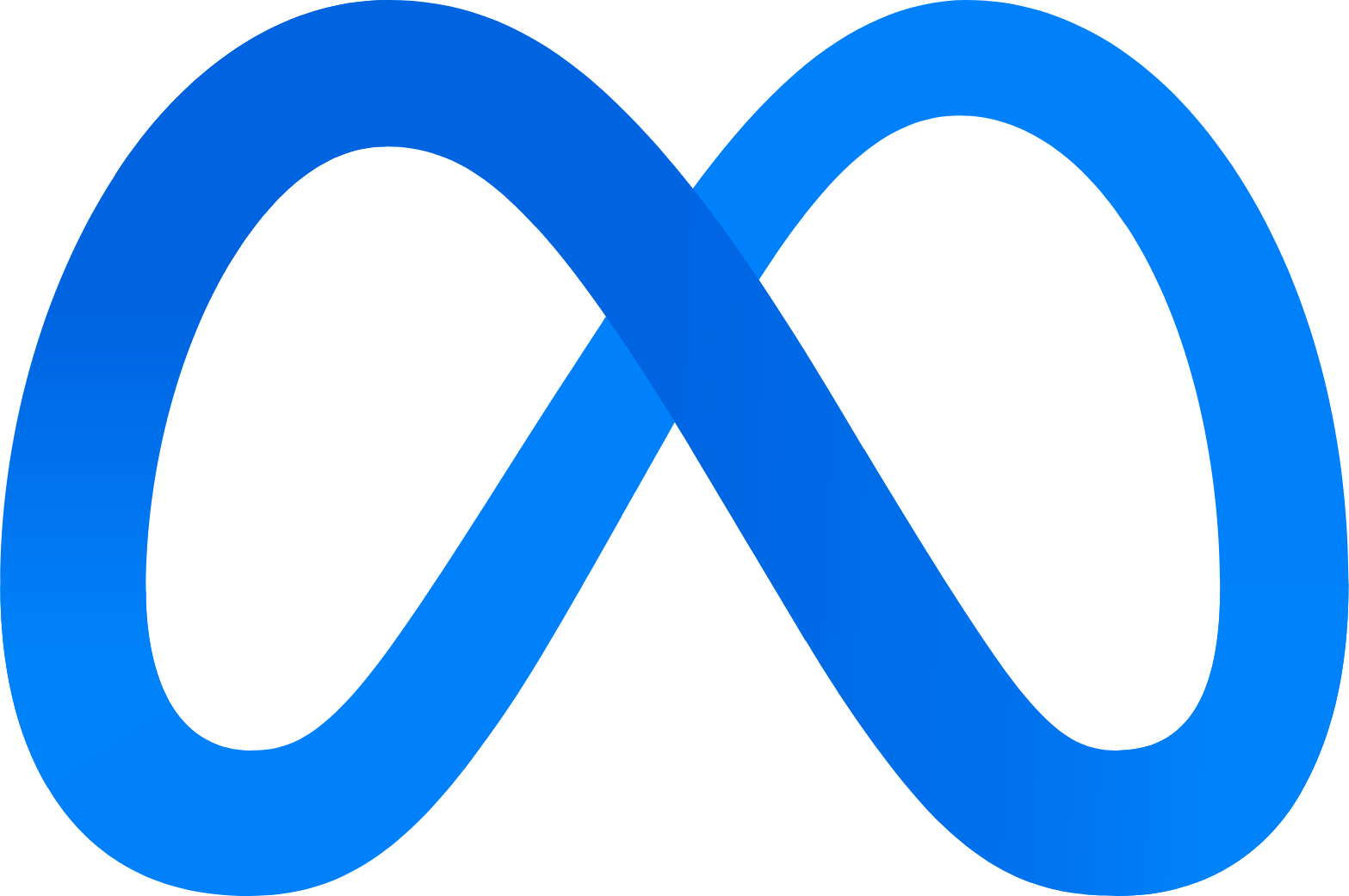} 
& sider 
& Automatic 
& Server-side 
& \redbox{PII}\hspace{0.5mm}\yellowbox{WC} 
& 4.35.0 \\

\textbf{Monica} - Your AI Copilot 
& 3M 
& \includegraphics[width=0.035\textwidth, height=0.02\textwidth]{figures/logos/ChatGPT-Logo.png}\includegraphics[width=0.017\textwidth, height=0.02\textwidth]{figures/logos/Google-Gemini-Logo.png} \includegraphics[width=0.02\textwidth, height=0.02\textwidth]{figures/logos/Claude-Logo.png} \includegraphics[width=0.022\textwidth, height=0.017\textwidth]{figures/logos/LLama-Logo.png} 
& gpt-4o-mini 
& Mixed 
& Server-side 
& \redbox{PII}\hspace{0.5mm}\yellowbox{UA}\hspace{0.5mm}\redbox{PC}\hspace{0.5mm}\redbox{FI} 
& 7.6.0 \\

\textbf{ChatGPT for Google} 
& 2M 
& \includegraphics[width=0.035\textwidth, height=0.02\textwidth]{figures/logos/ChatGPT-Logo.png}\includegraphics[width=0.017\textwidth, height=0.02\textwidth]{figures/logos/Google-Gemini-Logo.png} \includegraphics[width=0.02\textwidth, height=0.02\textwidth]{figures/logos/Claude-Logo.png} \includegraphics[width=0.022\textwidth, height=0.017\textwidth]{figures/logos/LLama-Logo.png} 
& gpt-4o-mini 
& Mixed 
& Client-side 
& \redbox{PII}\hspace{0.5mm}\yellowbox{UA}\hspace{0.5mm}\redbox{PC}\hspace{0.5mm}\redbox{FI} 
& 5.5.1 \\

\textbf{Merlin} Ask AI 
& 1M 
& \includegraphics[width=0.035\textwidth, height=0.02\textwidth]{figures/logos/ChatGPT-Logo.png}\includegraphics[width=0.017\textwidth, height=0.02\textwidth]{figures/logos/Google-Gemini-Logo.png} \includegraphics[width=0.02\textwidth, height=0.02\textwidth]{figures/logos/Claude-Logo.png} \includegraphics[width=0.022\textwidth, height=0.017\textwidth]{figures/logos/LLama-Logo.png} 
& gpt-4o 
& Mixed 
& Server-side 
& \redbox{PII}\hspace{0.5mm}\yellowbox{LOC} 
& 7.3.2 \\

\textbf{MaxAI}: Chat with Webpage 
& 900K 
& \includegraphics[width=0.035\textwidth, height=0.02\textwidth]{figures/logos/ChatGPT-Logo.png}\includegraphics[width=0.017\textwidth, height=0.02\textwidth]{figures/logos/Google-Gemini-Logo.png} \includegraphics[width=0.02\textwidth, height=0.02\textwidth]{figures/logos/Claude-Logo.png} \includegraphics[width=0.022\textwidth, height=0.017\textwidth]{figures/logos/LLama-Logo.png} 
& gpt-4o-mini 
& Manual 
& Server-side 
& \redbox{PII}\hspace{0.5mm}\yellowbox{UA} 
& 6.7.1 \\

\textbf{Perplexity} - AI Companion 
& 500K 
& \includegraphics[width=0.035\textwidth, height=0.02\textwidth]{figures/logos/ChatGPT-Logo.png}\includegraphics[width=0.02\textwidth, height=0.02\textwidth]{figures/logos/Claude-Logo.png} \includegraphics[width=0.022\textwidth, height=0.017\textwidth]{figures/logos/LLama-Logo.png} 
& perplexity 
& Manual 
& Server-side 
& \graybox{No Data} 
& 1.0.21 \\

\textbf{HARPA} AI 
& 400K 
& \includegraphics[width=0.035\textwidth, height=0.02\textwidth]{figures/logos/ChatGPT-Logo.png}\includegraphics[width=0.017\textwidth, height=0.02\textwidth]{figures/logos/Google-Gemini-Logo.png} \includegraphics[width=0.02\textwidth, height=0.02\textwidth]{figures/logos/Claude-Logo.png}
& harpa-v1-smart 
& Manual 
& Server-side 
& \redbox{PII}\hspace{0.5mm}\yellowbox{UA}\hspace{0.5mm}\redbox{WH}\hspace{0.5mm}\yellowbox{WC} 
& 9.6.2 \\

\textbf{TinaMind} - AI Assistant 
& 50K 
& \includegraphics[width=0.035\textwidth, height=0.02\textwidth]{figures/logos/ChatGPT-Logo.png}\includegraphics[width=0.017\textwidth, height=0.02\textwidth]{figures/logos/Google-Gemini-Logo.png} \includegraphics[width=0.02\textwidth, height=0.02\textwidth]{figures/logos/Claude-Logo.png}
& gemini-1.5-pro 
& Manual 
& Server-side 
& \redbox{PII}\hspace{0.5mm}\yellowbox{UA}\hspace{0.5mm}\redbox{PC} 
& 2.14.2 \\

\textbf{Copilot}: AI Assistant
& 30K 
& \includegraphics[width=0.035\textwidth, height=0.02\textwidth]{figures/logos/ChatGPT-Logo.png}\includegraphics[width=0.017\textwidth, height=0.02\textwidth]{figures/logos/Google-Gemini-Logo.png} \includegraphics[width=0.02\textwidth, height=0.02\textwidth]{figures/logos/Claude-Logo.png}
& gpt-4o-mini 
& Automatic 
& Server-side 
& \redbox{PII} 
& 1.5.73 \\

\bottomrule
\end{tabular}
\vspace{-6mm}
\end{table*}

\vspace{-3mm}
\section{Methodology}
\label{sec:audit-methodology}
\vspace{-2mm}

In this section, we introduce our auditing framework for GenAI browser assistants, designed to evaluate user tracking, profiling, and personalization, as shown in Figure~\ref{fig:methodology}. 
We make our crawling and analysis framework available at 
\url{https://doi.org/10.5281/zenodo.15618960}.

\subsection{Selection of GenAI Browser Assistants}
\label{sec:methodology-selecting-assistants}
\vspace{-1mm}

Among the three categories of GenAI browser assistants described in Section~\ref{sec:background}, we focus on \textit{search-based} assistants. 
This is due to several reasons: 
(1) Querying a search engine is relevant to querying an LLM model, allowing a natural integration of LLM-generated responses to a user's online searches. 
(2) It has been argued that search engines and LLMs would co-exist due to their unique individual capabilities to incorporate recent information and generate context-aware responses, respectively~\cite{Newman_Fletcher_Eddy_Robertson_Nielsen_2023}. 
(3) Although commonly referred to as ``search-based,'' these assistants' capabilities are not limited to search alone; they provide functionalities that allow their usage on any webpage as opposed to other categories of assistants that have limited scope and functionalities. 
(4) Combination of continued access to a user's complete browsing activity alongside GenAI capabilities raises privacy concerns, making them ideal candidates for our audit.

Our audit focuses on the most widely used web browser, Google Chrome.
We survey various extensions from the Chrome Web Store to identify ``AI-based search assistants" that offer the following three core capabilities. 
(1) direct integration with search engines (e.g., Google search) via a sidebar response; 
(2) interactive sidebar functionality to chat with any webpage; and 
(3) the ability to summarize webpage content.
We select nine of the most popular extensions that meet these criteria, based on installation numbers. 
Details about the selected assistants are provided in Table~\ref{tab:extension-details}.

\vspace{-3mm}
\subsection{Crawling Infrastructure}
\label{sec:methodology-crawling-infrastructure}
\vspace{-1mm}

\begin{figure*}
    \centering
    \includegraphics[width=1\linewidth]{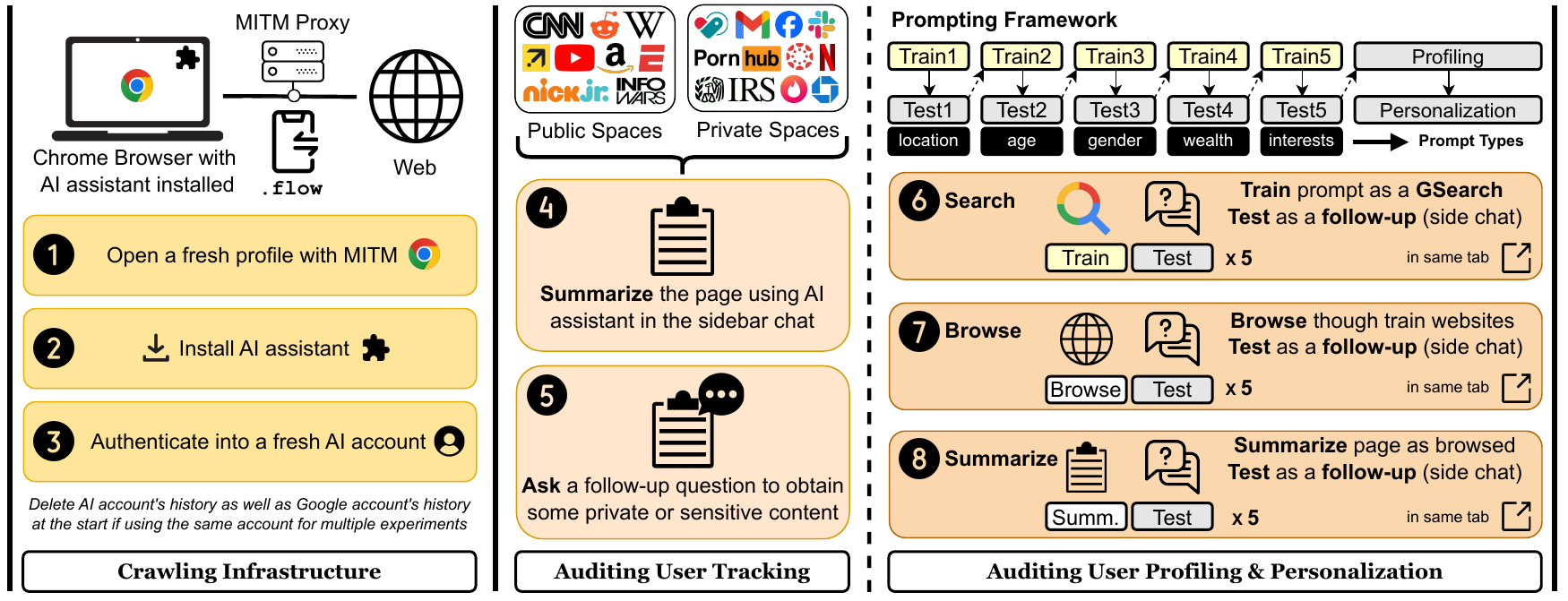}
    \vspace{-7mm}
    \caption{Experimental design of our auditing framework for AI browser assistants.}
    \label{fig:methodology}
    \vspace{-6mm}
\end{figure*}

To audit data collection and tracking practices of GenAI browser assistants, it is important to examine network traffic activity generated by the browser with the extension installed. 
We use a dedicated laptop to conduct our experiments. 
The laptop is connected to the Internet via a Wi-Fi network and has Chrome browser installed.
To intercept and log decrypted network traffic while browsing, we install and configure \texttt{Mitmproxy}~\cite{mitmproxy} -- an open-source HTTPS proxy -- on the device following the official guidelines. 
During the proxy configuration phase, \texttt{Mitmproxy} certificate is imported into Chrome's Trusted Root Certification Authorities, allowing \texttt{Mitmproxy} to intercept and decrypt HTTPS traffic without triggering security warnings.
The proxy server acts as a man-in-the-middle between the laptop and the Internet, as shown in Figure~\ref{fig:methodology}.
This allows us to capture all incoming and outgoing real-time web traffic while conducting our experiments, including requests, responses, payloads, and browser storage.

To conduct the experiments described in Sections \ref{sec:methodology-audit-user-tracking} and \ref{sec:methodology-audit-profiling-personalization}, we initialize a new \texttt{Mitmproxy} instance on localhost with \texttt{mitmweb} using command line interface (CLI).
Next, we launch Chrome browser by specifying a new custom directory for user data to open a fresh browser instance with a new profile each time (Step {\large \ding{182}}).
This ensures a clean browsing session with no past history associated with the profile, 
providing independence across experiments. 
Next, we visit the Chrome web store, search for the browser assistant, and install its extension into the browser instance (Step {\large \ding{183}}). 
All assistants in our study, except Perplexity, require a user login. 
Perplexity can be used with or without logging-into an account.
To keep the profiles isolated across experiments, we sign-up for a new user account each time using a distinct temporary email provided by various online services (Step {\large \ding{184}}). 
However, some assistants do not allow the use of a temporary email address and instead require authentication via Google or Apple accounts. 
In such cases, we reuse a designated Google account, since Google restricts the number of accounts tied to a single phone number.
To avoid contamination across different experiments for a given extension, before the start of each experiment, we manually open a fresh browser instance and delete all Google activity and browsing history. 
We also delete the assistant's chat history and any account-associated memory.
We acknowledge that the browser assistant may store a server-side mapping of the user's account-related activity -- which may or may not be deleted when the user performs client-side deletion. 
This is a limitation of our methodology.
After setup, we conduct the experiments described in the following sections.
At the end of each experiment, web traffic from the entire browsing session is stored in a \texttt{.flow} file. 
We describe analysis of \texttt{.flow} files in Section~\ref{sec:methodology-network-traffic-analysis}.

\vspace{-4mm}
\subsection{Auditing User Tracking}
\label{sec:methodology-audit-user-tracking}

\vspace{-2mm}
When navigating through the web with an installed GenAI browser assistant, a user may visit a mix of public and private online spaces. 
In these spaces, a user may use the assistant without realizing the consequences of doing so. 
For instance, a user browsing through their health records may inadvertently share sensitive information due to the use of the browser assistant. 
It is important to understand if the assistants have built-in safeguards to prevent collection and exfiltration of user's sensitive information or do they freely collect and share such information at all times.

To understand such implicit data collection and sharing, we identify 20 commonly-browsed website categories: 10 public spaces and 10 private spaces as listed in Table~\ref{tab:content_categories}.
For each category, we select one popular website to experiment with. 
Websites classified as private require user authentication and comprises personal or sensitive information that a user would not want to be collected or shared.
We test each website with every extension, resulting in a total of 180 experiments (20 websites x 9 extensions).
Using our crawling infrastructure, we visit the website after logging-into the browser assistant's account. 
Next, if it is a private space website, we log in to the test website using (our own) personal account.
If applicable, we navigate through the website by clicking on specific links that lead us to the webpage containing private information about the user.
For example, we navigate to `Medical Records' tab and click on a specific visit to open a detailed medical record associated with some visit. 
To avoid inconsistencies and ensure comparability, we always perform the same set of actions (if any) for a given website across different experiments.
Once the website is loaded and navigations are complete, we invoke the sidebar interface and perform summarization of the webpage using its `summary' feature (Step {\large \ding{185}}).
Finally, we ask a follow-up questions (see Table~\ref{tab:content_categories} in appendix) to explicitly request information displayed on the webpage (Step {\large \ding{186}}).
This is to understand whether the assistant recognizes the page content to be sensitive, problematic, copyrighted, or personal to avoid providing the requested information or not. 
Responses from the browser assistants as well as \texttt{.flow} files are stored for further analysis.

\vspace{-3mm}
\subsection{Auditing Profiling and Personalization}
\label{sec:methodology-audit-profiling-personalization}
Profiling refers to the process of collecting user data to infer attributes such as demographics, preferences, or interests.
Personalization builds on profiling by using these inferences to tailor responses or experiences to the individual user.
It is important to understand whether GenAI browser assistants associate the collected information with the user's profile to answer specific questions or personalize their responses

\noindent \textbf{{Prompting Framework}.} 
To evaluate profiling and personalization in GenAI browser assistants with respect to RQ3, we propose a novel prompting framework as depicted in Figure~\ref{fig:methodology}. 
The profile we develop represents -- ``\textit{A rich millennial male from southern California who is interested in equestrian activities}''.
We formulate 5 Train-Test prompt pairs -- each explicitly leaking some attribute about the user in the train prompt and then testing for profiling based on the leaked information via a test prompt. 
The user's attribute that we leak and test for in chronological order of prompts include location, age, gender, wealth, and interest. 
After the last test prompt, we re-ask all the 5 test prompts combined together as a single \textit{profiling prompt} for the GenAI browser assistant. 
This is followed by ``personalization prompt'' where we simply ask the assistant to suggest Top 3 activities that the user would likely include in their itinerary of a vacation, based on what it has learnt about the user.
The profiling and personalization prompts are first asked in the same tab where rest of the training or testing of individual attributes occur to understand its \textit{in-context} behaviour.
Next, we also open a new browser tab within the same session and re-ask them to understand its \textit{out-of-context} behaviour.
Except personalization and train prompts, we condition all responses to 5 test prompts and the profiling prompt to only output a binary response -- "Yes" or "No" to avoid any subjectivity.
This is achieved by adding a meta (system) prompt.
Personalization prompt is reported to personalize responses (i.e., "Yes") if one of the suggested activities clearly lists atleast one `equestrian activity' (i.e., the user's leaked interest).

\noindent \textbf{{Selection and Ordering of Prompts}.} 
We explicitly choose a niche interest for our profile to avoid matches with generic activities that an LLM model might otherwise suggest such as `hiking' for instance.
This minimizes the risk of spurious correlations between the leaked interest and the personalized responses.
User descriptions of each prompt were crafted in accordance with the attribute we wanted to test for. 
To test for a new attribute, future work can easily generate a description that leaks the attribute by asking an LLM model and providing our prompts as a context.
Additionally, each train prompt targets a distinct user attribute (e.g., location, age). 
Thus, we do not expect any meaningful difference from either the specific ordering or the prompt used.
Section~\ref{sec:prompting} in appendix lists all prompts used in this study.

\noindent \textbf{{Experimentation}.} 
Amongst different features provided by search-based GenAI browser assistants, we specifically test three most useful features that were observed to be common across all 9 extensions -- search-based integration, webpage chat, and webpage summary. 
These are studied under the following four scenarios for each extension (totaling to 36 experiments) as described in Figure~\ref{fig:methodology}:

\noindent \textbf{{Control}.} 
In control, we simply ask each test prompt (without train prompts) followed by profiling and personalization prompts -- each in a new tab.

\noindent \textbf{{\large \ding{187}} {Search}.} 
In search scenario, train prompts are entered into Google search as a normal search query. 
When Google search is performed, it may result in an automatic response generation from the GenAI assistant, which is displayed to the user as a sidebar response. 
In cases where automatic responses are not generated, the train prompt is explicitly asked to the GenAI assistant in the sidebar chat. 
Next, the test prompt corresponding to a train prompt is asked as a follow-up question in the sidebar chat.
The same steps are repeated for each train-test prompt pair within the same browser tab.
At the end, profiling and personalization prompts are asked both -- in-context and out-of-context.

\noindent \textbf{{\large \ding{188}} {Browse}.} 
In this scenario, no interaction occurs during the training phase. Training phase involves browsing through 10 webpages -- 2 pages per leaked attribute. 
These webpages (listed in Table~\ref{Ablation_URLs} in appendix) are selected from the top-ranking Google search results of the specific attribute descriptions leaked in the finalized prompts. 
We manually confirmed that the website content explicitly reveals the given attribute. 
Other than scrolling a page to explore its content, we also click on internal links to emulate normal browsing.
The set of clicks are fixed and remains the same across different experiments involving the same website.
Our hypothesis is that if GenAI browser assistant collects information about the user's visit to different webpages, then it may use it to infer user's attributes.
The experiment ends by asking 5 test prompts along with profiling and personalization prompts sequentially within the same tab in the sidebar chat. Profiling and personalization prompt are also repeated out-of-context.

\noindent \textbf{{\large \ding{189}} {Summarize}.} 
We use the \textit{summarize webpage} feature of GenAI browser assistants to understand if usage of this feature aids the browser assistant in building a profile about the user and later on use it to personalize its responses. 
We visit the same 10 webpages as before, performing the same interactions as in \textit{Browse}.
We also summarize each page.
In this scenario, summarization acts as training.
Finally, the same prompting as in \textit{Browse} is performed.

We repeat each prompting experiment in our study 15 times and report the majority response to account for variability in the output due to probabilistic nature of GenAI systems.

\vspace{-5mm}
\subsection{Semi-automated Analysis}
\label{sec:methodology-network-traffic-analysis}

In this section, we discuss our approach to perform network traffic analysis based on the web traffic data collected in the form of \texttt{.flow} files during the experiments.
We analyze the \texttt{.flow} files corresponding to the majority response across repeated runs.
We use Python module \texttt{mitmproxy.io} to programmatically parse the network flows and extract details about the contacted endpoints, request and response headers, payloads, and responses. 
We identify each request to be first-party if its domain matches the domain of the browser assistant's website, and third-party otherwise.
To avoid any misclassifications due to organizational relatedness of different domains, we further use DuckDuckGo's entity list that maps domains to their parent organization.
If two domains were found to belong to the same organization, and one was initially classified as first-party while the other as third-party, we reclassified the latter as first-party.
A flow is also classified as either a foreground flow or a background flow. Background flows represent extension traffic emerging from the background service worker of the browser assistant. 
They are identified based on the value of a request header -- \texttt{origin}.
A value of the form \url{chrome-extension://<extension-id>} represents traffic from an extension, where \texttt{extension-id} can be obtained from extension's page on Chrome web store.
An automated script extracts the flows from different experiments related to a given extension along with the above details in the form of a CSV file. 
We perform detailed analysis of these files to extract user, device, webpage, or browsing related information from the payload.


\vspace{-3mm}
\section{Results}
\label{sec:results}

\vspace{-1mm}
\subsection{Architecture of GenAI browser assistants}
\label{sec:results-architecture}

This section discusses our findings related to RQ1, providing a deeper understanding of how GenAI browser assistants are designed. 
We qualitatively analyze network traffic activity to learn the following architectural differences across 9 assistants: (1) backend model, (2) response architecture, (3) context restrictions, and (4) response variability.

\noindent \textbf{{Backend model}.}
Since GenAI browser assistants function as wrappers on top of the proprietary LLM models, we observe that they provide support for most of the popular models -- OpenAI's chatGPT~\cite{chatgpt2025}, Google's Gemini~\cite{gemini2025}, Anthropic's Claude AI~\cite{Anthropic}, and Meta's Llama~\cite{llama2025} as shown in Table~\ref{tab:extension-details}. 
For each extension, the request to fetch the response also contained the model information. 
We observed \texttt{gpt-4o-mini} to be the most popularly used default model due to its cost-efficiency trade-off for assistant developers as well as for users over other models.
Some extensions used different models for different user activities. For example, ChatGPT for Google (CFG) used \texttt{gpt-4o-mini} for Google-search-triggered automatic response generation while it used \texttt{gpt-3.5-turbo} when a user explicitly opened a sidebar chat to ask some question. 
Sider, Perplexity, and Harpa use custom LLM models, making it difficult to determine how their designs differ from commonly used proprietary models.

\noindent 

\noindent \textbf{{Response architecture}.}
Browser assistants may contact popular proprietary GenAI models to generate responses either from client-side or server-side. Figure~\ref{fig:threat-model} depicts the most common case we observed in our study, where 8 out of 9 extensions operate server-side. 
This means that first, the user's selected model (or default in case of no selection) is shared along with the user's query and other metadata to the browser assistant's server. 
Based on the selected model, the assistant's server calls the corresponding LLM API to direct the user's request to. 
The personalized response is finally displayed to the user.
In case of CFG, response generation is initiated from the client-side (see Table~\ref{tab:extension-details}). 
To use CFG browser assistant, the user needs to link their OpenAI's \url{chatgpt.com} account with CFG. 
As a result, all queries from interaction with CFG are directly shared with \url{chatgpt.com} to fetch the response. 
We also analyze if browser assistants always monitor user's activities (passively) or if they perform monitoring only when user explicitly interacts with the assistant (actively).
Here, we refer to browser activities such as scrolling, clicks, navigation, etc. that is not tied to explicit assistant features. 
We found it to be always ``active'' for all 9 extensions, suggesting that they do not passively monitor users' non-extension-related activities.
In regards to extension-related activities like search, we found that some assistants such as Sider and Copilot auto-invoke response generation when a user submits a search query in the search engine. 4/9 extensions require manual invocation of the assistant to obtain a response while remaining 3 work in a mixed fashion.

\noindent \textbf{{Context restrictions}.}
Next, we compare the interdependence of context used by the browser assistant under different scenarios. 
We look at two sets of context -- sidebar response versus popup sidechat and context across page navigations. 
In the former case, during our experiments, we analyze dependency of context from the response generated by the assistant in the empty sidebar space of Google search results page with the context of the sidebar chat that pops up upon clicking the extension icon. 
We conducted a simple experiment in which we searched `I like apples' on Google Search (Q1) and ensured that the sidebar response shows up. 
Next, we opened the sidebar popup chat and asked `Do I like apples?' (Q2) -- if it responds `Yes' then that suggests the two contexts are dependent, otherwise independent. 
We observed that 4 assistants -- ChatGPT for Google, Max AI, TinaMind, and Copilot maintained the same context while the remaining 5 maintained separate contexts, suggesting stronger privacy. 
In the second case, we ask Q1 in the sidebar popup chat while visiting a specific webpage; next we navigate to a different webpage and ask Q2 to determine whether the assistant preserved context. 
We observe that context was not reset only for one assistant -- ChatGPT for Google, suggesting its capability to remember contexts across multiple websites as user browses through the web. 
We found Perplexity to be the most private assistant as it explicitly displayed the message stating `\textit{I do not have the ability to recall previous interactions or questions. Each session or question is treated independently for privacy reasons.}'.
We further analyze what types of data form the necessary context for different assistants in more detail in Section~\ref{sec:results-tracking}.

\noindent \textbf{{Response Variability}.}
It is important to understand that even when different browser assistants use the same model to process a user's query, the output may vary significantly. 
Several factors play a role in this. We discuss two of the most important ones below: \textit{Temperature} and \textit{System prompts}.

\noindent \textbf{{Temperature}.}
First, as aforementioned, output of generative models is probabilistic and is influenced by the \texttt{temperature} parameter.
%
For example, OpenAI's ChatGPT provides the capability to set \texttt{temperature} in the range of 0 to 1, where lower values provide more deterministic and focused outputs while higher values generate more creative or random responses~\cite{OpenAI}.
Different browser assistants may set and use different values of \texttt{temperature}, resulting in differences in outputs even when they use the same model.
Moreover, even the same browser assistant may dynamically adjust \texttt{temperature} based on the nature of the user query.
For example, when asked to a browser assistant: ``What do you know about me?'' -- it may query OpenAI's API with a higher temperature value to avoid generating a precise response and rather keep it open-ended. 
However, if the same user asks the same browser assistant: ``What is the distance between New York and Washington D.C.?'', the assistant may use a smaller value of \texttt{temperature} to provide a deterministic output.

\noindent \textbf{{System Prompts}.}
Second, system prompts are provided to the LLM models along with the user query for various reasons such as to enhance task-specific accuracy, provide contextual guidance, prevent unwanted output, ensure output formatting, and mitigate vulnerabilities such as jailbreaking~\cite{wallace2024instruction}.
During the network traffic analysis, we discovered various system prompts being employed by different browser assistants as listed in Section~\ref{system_prompts_appendix} of the appendix. 
We observe that some assistants use stricter system prompts to answer a user's query more accurately than others.
For example, Sider's system prompt to answer a question simply states to ``\textit{use simple and clear language}'' whereas Harpa's system prompt explicitly states to ``\textit{NEVER fabricate, infer, or guess information. Do not hallucinate links. Be to the point}''. 
In the same prompt, it can be seen that Harpa includes \texttt{$\{\{user\_info\}\}$} as context
\renewcommand{\arraystretch}{1.2} 

\definecolor{CBRed}{HTML}{D55E00}    
\definecolor{CBOrange}{HTML}{E69F00} 
\definecolor{CBYellow}{HTML}{F0E442} 
\definecolor{CBGreen}{HTML}{009E73}  

\newcommand{\greencell}{\cellcolor{CBGreen!75}}
\newcommand{\yellowcell}{\cellcolor{CBYellow!75}}
\newcommand{\orangecell}{\cellcolor{CBOrange!75}}
\newcommand{\redcell}{\cellcolor{CBRed!75}}

\newcommand{\boldX}{\tikz[baseline]{\node[draw=none, text=black, align=center, font=\bfseries] {X};}}
\newcommand{\boldSkip}{\tikz[baseline]{\node[draw=none, circle, text=black, align=center, font=\bfseries, thick] {-};}}
\newcommand{\boldTick}{\tikz[baseline]{\node[draw=none, text=black, align=center, font=\bfseries, thick] {\checkmark};}}
\newcommand{\boldTickSlash}{\tikz[baseline]{\node[draw=none, text=black, align=center, font=\bfseries, thick] {\checkmark}; \draw (-0.1,0.1) -- (0.2,0.5);}}
\newpage

{\setlength{\extrarowheight}{1pt}
\begin{table*}[!ht]
\caption{Data collection and exfiltration behavior of assistants in public and private online spaces of a user. \textit{Exfiltration legend}:} \vspace{-2.5mm}\protect\Exfilredbox{Full Webpage}: Page text, title, location, hyperlinks. \protect\Exfilyellowbox{Server-fetch Webpage}: Page title, location, server-fetched file's upload location. \protect\Exfilorangebox{Plain Webpage}: Page text, title, location. \protect\Exfilgreenbox{Partial Webpage}: Partial content or missing details. \textit{Response legend}: \\ \vspace{0.35mm} \ding{52}: Response with relevant details. \HalfCheckmark: Missing some details in response \RestrictionSymbol: Response restricted. \ding{56}: No response generated.
\label{tab:data-collection}
\normalsize



\centering
\renewcommand\arraystretch{1} 
\setlength{\tabcolsep}{4pt} 
\begin{adjustbox}{max width=\textwidth}
\begin{tabular}{|m{.4cm}|m{3.8cm}|m{3.3cm}|*{9}{>{\centering\arraybackslash}m{.75cm}|}}
\hline
 \normalsize & \makecell[c]{\hspace{-.5em}\vspace{-2em}Category\vspace{0.5em}\hspace{-0.2em}} 
 &  \makecell[c]{\hspace{-0.25em}\vspace{-2em}WebPage\vspace{0.5em}\hspace{-0.2em}}  &
\makecell[c]{\raisebox{-1.1em}{\rotatebox{90}{Sider~~}}} &
\makecell[c]{\raisebox{-2em}{\rotatebox{90}{Monica~~}}} &
\makecell[c]{\raisebox{-1.1em}{\rotatebox{90}{CFG~~~}}} &
\makecell[c]{\raisebox{-1.6em}{\rotatebox{90}{Merlin~~}}} &
\makecell[c]{\raisebox{-1.9em}{\rotatebox{90}{MaxAI~~}}} &
\makecell[c]{\raisebox{-.7em}{\rotatebox{90}{~Perplexity~}}} &
\makecell[c]{\raisebox{-1.5em}{\rotatebox{90}{Harpa~~~}}} &
\makecell[c]{\raisebox{-2.1em}{\rotatebox{90}{~TinaMind~}}} &
\makecell[c]{\raisebox{-1.8em}{\rotatebox{90}{Copilot~~~}}} \\
\hline
\renewcommand\arraystretch{1.5}
\multirow{9}{*}{\rotatebox{90}{Public Spaces}} 
& News Platforms & cnn.com & \redcell\ding{52} & \orangecell\ding{52} & \orangecell\ding{52} & \orangecell\ding{52} & \orangecell\ding{52} & \yellowcell\ding{56} & \redcell\ding{52} & \greencell\ding{52} & \greencell\ding{56} \\ \cline{2-12}

& Open Forums & reddit.com & \greencell\HalfCheckmark & \orangecell\ding{52} & \orangecell\HalfCheckmark & \orangecell\ding{52} & \orangecell\HalfCheckmark & \yellowcell\HalfCheckmark & \redcell\ding{52} & \greencell\HalfCheckmark & \greencell\ding{52} \\ \cline{2-12}

& Informative Articles & wikipedia.org & \greencell\ding{52} & \redcell\ding{52} & \orangecell\ding{52} & \orangecell\ding{52} & \orangecell\ding{52} & \yellowcell\ding{52} & \redcell\ding{52} & \greencell\ding{52} & \greencell\ding{56} \\ \cline{2-12}

& E-commerce Website & amazon.com & \greencell\ding{52} & \orangecell\ding{52} & \orangecell\ding{52} & \orangecell\ding{52} & \orangecell\HalfCheckmark & \yellowcell\ding{56} & \redcell\ding{52} & \greencell\HalfCheckmark & \greencell\ding{56} \\ \cline{2-12}

& Sports Websites & espn.com & \redcell\ding{52} & \redcell\ding{52} & \orangecell\ding{52} & \orangecell\ding{52} & \orangecell\ding{52} & \yellowcell\ding{52} & \redcell\ding{52} & \greencell\HalfCheckmark & \greencell\ding{56} \\ \cline{2-12}

& Travel Platforms & expedia.com & \greencell\ding{52} & \redcell\ding{52} & \orangecell\ding{52} & \orangecell\ding{52} & \orangecell\ding{52} & \yellowcell\ding{56} & \redcell\ding{52} & \greencell\ding{52} & \greencell\ding{52} \\ \cline{2-12}

& User-generated Media & youtube.com & \greencell\HalfCheckmark & \greencell\ding{56} & \redcell\ding{52} & \orangecell\ding{52} & \orangecell\HalfCheckmark & \yellowcell\ding{56} & \redcell\HalfCheckmark & \greencell\ding{52} & \greencell\ding{52} \\ \cline{2-12}

& Kids Website & nickjr.com & \greencell\ding{56} & \orangecell\ding{52} & \orangecell\ding{52} & \orangecell\ding{52} & \orangecell\ding{52} & \yellowcell\ding{52} & \redcell\ding{52} & \greencell\ding{52} & \greencell\ding{56} \\ \cline{2-12}

& Misinformation Website & infowars.com & \redcell\ding{52} & \redcell\ding{52} & \redcell\ding{52} & \orangecell\ding{52} & \orangecell\ding{52} & \yellowcell\ding{56} & \redcell\ding{52} & \greencell\ding{52} & \greencell\ding{52} \\ \cline{2-12}

& Violence Material & guns.com & \redcell\ding{52} & \orangecell\ding{52} & \orangecell\ding{52} & \orangecell\ding{52} & \orangecell\ding{52} & \yellowcell\ding{56} & \redcell\ding{52} & \greencell\ding{52} & \greencell\ding{56} \\ 

\hline
\multirow{10}{*}{\rotatebox{90}{Private Spaces}} 
& Health Portal & hem.ucdavis.edu & \redcell\ding{52} & \orangecell\ding{52} & \orangecell\ding{52} & \redcell\ding{52} & \orangecell\ding{52} & \yellowcell\ding{56} & \redcell\HalfCheckmark & \greencell\HalfCheckmark & \greencell\ding{56} \\ \cline{2-12}

& Email Account & mail.google.com & \redcell\ding{52} & \redcell\ding{52} & \redcell\HalfCheckmark & \orangecell\ding{52} & \orangecell\HalfCheckmark & \yellowcell\ding{56} & \redcell\HalfCheckmark & \greencell\ding{52} & \greencell\ding{56} \\ \cline{2-12}

& Social~Media~Platform & facebook.com & \redcell\ding{52} & \orangecell\ding{52} & \orangecell\ding{52} & \orangecell\ding{52} & \orangecell\ding{52} & \yellowcell\ding{56} & \redcell\ding{52} & \greencell\HalfCheckmark & \greencell\ding{56} \\ \cline{2-12}

& Adult Content & pornhub.com & \greencell\HalfCheckmark & \orangecell\HalfCheckmark & \orangecell\HalfCheckmark & \orangecell\HalfCheckmark & \orangecell\HalfCheckmark & \yellowcell\RestrictionSymbol & \redcell\HalfCheckmark & \greencell\HalfCheckmark & \greencell\ding{56} \\ \cline{2-12}

& Online~Streaming~Service & netflix.com & \redcell\HalfCheckmark & \orangecell\HalfCheckmark & \orangecell\HalfCheckmark & \orangecell\HalfCheckmark & \orangecell\HalfCheckmark & \yellowcell\ding{56} & \redcell\HalfCheckmark & \greencell\HalfCheckmark & \greencell\ding{56} \\ \cline{2-12}

& Government Website & irs.gov & \orangecell\HalfCheckmark & \orangecell\HalfCheckmark & \orangecell\HalfCheckmark & \redcell\HalfCheckmark & \orangecell\HalfCheckmark & \yellowcell\ding{56} & \redcell\ding{52} & \greencell\HalfCheckmark & \greencell\ding{56} \\ \cline{2-12}

& Dating Service & tinder.com & \greencell\ding{56} & \orangecell\ding{52} & \orangecell\HalfCheckmark & \redcell\HalfCheckmark & \orangecell\HalfCheckmark & \yellowcell\ding{56} & \redcell\ding{52} & \greencell\HalfCheckmark & \greencell\ding{56} \\ \cline{2-12}

& Financial Service & chase.com & \greencell\ding{56} & \orangecell\HalfCheckmark & \redcell\HalfCheckmark & \redcell\HalfCheckmark & \orangecell\HalfCheckmark & \yellowcell\ding{56} & \redcell\HalfCheckmark & \greencell\ding{56} & \greencell\ding{56} \\ \cline{2-12}

& Educational Platform & canvas.instructure.com & \redcell\ding{52} & \orangecell\ding{52} & \redcell\ding{52} & \redcell\ding{52} & \orangecell\ding{52} & \yellowcell\ding{56} & \redcell\ding{52} & \greencell\HalfCheckmark & \greencell\ding{52} \\ \cline{2-12}

& Messaging Platform & slack.com & \redcell\ding{52} & \redcell\ding{52} & \redcell\ding{52} & \orangecell\ding{52} & \orangecell\ding{52} & \yellowcell\ding{56} & \redcell\ding{52} & \greencell\ding{52} & \greencell\ding{56} \\

\hline
\end{tabular}
\end{adjustbox}
\vspace{-6mm}
\end{table*}}

\noindent along with the user's query to generate a response, suggesting that it shares user details with proprietary LLM models freely with each query. Moreover, Harpa's system prompt states ``\textit{Please ignore all previous instructions}'' to prevent any jail-breaking attempts through cleverly-crafted user queries. 
TinaMind's system prompts include user-specific metadata such as language preference, timestamp, and timezone details with each query in order to interpret the user's location-dependent queries as per the applicable region to produce a more relevant response.

\vspace{-5mm}
\subsection{User tracking by GenAI browser assistants}
\label{sec:results-tracking}
To systematically audit user tracking by different browser assistants, in Section~\ref{sec:results-implicit-tracking}, we first perform analysis of implicit collection and sharing of user's data as user visits public and private spaces online. 
We refer to it as `implicit' as the user does not actually intend to share their personal information with the GenAI assistant. 
However, the user may inadvertently let the assistant access it. 
Here, we analyze tracking of user data in context of the webpage content that the user is visiting. 
Next, in Section~\ref{sec:results-explicit}, we evaluate explicit data collection and sharing when using different functionalities offered by assistants, where we use the prompting framework to explicitly leak user attributes and understand tracking of all kinds of data associated with the user. 
Lastly, in Section~\ref{sec:privacy-policy-analysis}, we qualitatively compare the observed data collection and tracking practices against the privacy policy of the assistants.

\vspace{-4mm}
\subsubsection{Implicit collection and sharing of user data}
\label{sec:results-implicit-tracking}

In this section, we discuss the results related to implicit user tracking to see if and when do browser assistants collect data.
We followed the methodology described in Section~\ref{sec:methodology-audit-user-tracking} to visit different websites in public and private spaces, summarize the page content, and ask a follow-up question. 
Table~\ref{tab:data-collection} showcases results from running a total of 180 experiments across 20 websites and 9 assistants. 

%
The primary goal of the analysis is to understand what kind of webpages are vulnerable to collection of page content related data by browser assistants. We particularly focus on page content due to its direct impact on user's privacy. This is because if the visited webpages happen to be one of the private spaces to the user, then it puts their private data at a risk of getting collected. Besides user privacy, data collection of problematic content can result in browser assistant providing harmful responses to the user, while collection of copyrighted content may have regulatory implications. 

As part of the page data, we consider page title, page text, page location and any embedded links on the page. The cells in orange red suggests most egregious data collection, where complete webpage details were collected, while cells in green correspond to inefficient collection of pages (e.g., with missing details). We observe that Harpa collects full DOM in all 20 online spaces, while TinaMind is unable to extract full webpage details across all the experiments. This is likely due to two reasons -- one, TinaMind only focuses at capturing data that is in the viewport of the user, failing to capture rest of the page and two, its page extraction mechanism is less aggressive. The latter inference is based on the fact that it performed incomplete extraction even from the user's viewport content, often failing to reproduce the numerical content present on the webpage in its responses. 


Browser assistants can easily employ practices that can allow them to determine if a page contains sensitive or private data or not. For instance, when summarize feature is used with Perplexity, it shares the page URL with its own server and performs a server-side fetch of the webpage. This is inferred based on the file upload location received in response to the shared page location. Therefore, if the webpage belongs to a user's private authenticated space, Perplexity's server will never have access to the user's personal data. However, it will still be able to provide responses for public spaces. 

Merlin, MaxAI, ChatGPT for Google and Monica were all able to extract webpage contents for all 20 scenarios. 
Merlin was discovered to be the only assistant that recorded even the contents of the forms on webpages as opposed to all other browser assistants -- that did not collect form data. 
For example, it was able to collect the user's \textit{Social Security Number (SSN)} entered in a form on IRS refund portal. 
Email addresses as well as the full email thread were also collected by these assistants. 
One of the most surprising findings was that GenAI browser assistants collected and shared data to their own servers on authenticated patient portals. 
They were able to answer follow-up questions ranging from patient details to their medical history. 
Collection of such sensitive information without opt-in user consent is a potential violation of HIPAA~\cite{(OCR)_2024}. 
Moreover, student academic records including assessment scores, exam performances, overall grades -- were all collected and shared with browser assistant's servers, which is a potential violation of FERPA~\cite{FERPA}.

Overall, it can be observed that responses with missing details are more prominent in the lower half of the table suggesting that private spaces are handled with care by most extensions. However, this is insufficient since they are still collecting partial data that is private to the users. Adult content, online streaming services, dating platforms, and financial services -- were unanimously responded inefficiently by all the assistants. Titles displayed on Netflix homepage, preferences set on Tinder and transaction amounts as well as last four digits of card number on Chase were either inaccurately output or were missing from the output of different assistants. This is because we observed that the assistants missed capturing \texttt{shadow} elements from DOM of the webpage. On the brighter side, Perplexity actively suppressed response generation on adult pages due to explicit content stating ``\textit{I apologize, but I cannot provide information about or analyze pornographic content. This type of material is not appropriate for me to discuss. Perhaps I could assist you with a different, non-explicit topic instead?}''. However, Perplexity still recommended links to pornographic content in follow-up suggestions, showing inefficiencies in their model.

\subsubsection{Explicit collection and sharing of user data }
\label{sec:results-explicit}
\vspace{-2mm}

Having looked at what online spaces are vulnerable to data collection, we now focus on understanding what attributes are explicitly collected about the user, device, or page when different features provided by the browser assistant are used by the user -- namely, search, browse and summarize as shown in Table~\ref{tab:data-collection-sharing}.
\begin{table}[h!]
    \centering
    \renewcommand{\arraystretch}{1.2} 
    \setlength{\tabcolsep}{2.0pt} 
    \caption{Data sharing across extensions. \label{tab:data-collection-sharing} \textit{Legend: } \protect$\protect\circleLeft$ = First-party sharing, \protect$\protect\circleRight$ = Third-party sharing, \protect$\protect\circleFull$ = Both, \protect$\protect\circleNone$ = None.}
    \begin{tabular}{>{\centering\arraybackslash}m{0.3cm} p{1.8cm} *{14}{>{\centering\arraybackslash}p{0.3cm}}} 
        & \textbf{Scenario} 
        & \rotatebox{90}{Page Location} 
        & \rotatebox{90}{Page Content} 
        & \rotatebox{90}{GSearch Results} 
        & \rotatebox{90}{User's Query} 
        & \rotatebox{90}{Chat ID} 
        & \rotatebox{90}{Chat History} 
        & \rotatebox{90}{User Details} 
        & \rotatebox{90}{Device Details} 
        & \rotatebox{90}{Query Timestamp} 
        & \rotatebox{90}{Timezone} 
        & \rotatebox{90}{Referrer} 
        & \rotatebox{90}{Cookies} 
        & \rotatebox{90}{User Agent}
        & \rotatebox{90}{Local Storage} \\ 
        \midrule
        \multirow{3}{*}{\rotatebox{90}{\textbf{Sider}}}
        & Search & \circleLeft & \circleLeft & \circleLeft & \circleLeft & \circleLeft & \circleNone & \circleFull & \circleLeft & \circleLeft & \circleLeft & \circleNone & \circleLeft & \circleNone & \circleNone \\
        & Browse & \circleLeft & \circleLeft & \circleNone & \circleLeft & \circleLeft & \circleNone & \circleFull & \circleLeft & \circleLeft & \circleLeft & \circleNone & \circleLeft & \circleNone & \circleNone \\
        & Summarize & \circleLeft & \circleLeft & \circleNone & \circleLeft & \circleLeft & \circleNone & \circleFull & \circleLeft & \circleLeft & \circleLeft & \circleNone & \circleLeft & \circleNone & \circleNone \\
        \midrule
        \multirow{3}{*}{\rotatebox{90}{\textbf{Monica}}}
        & Search & \circleLeft & \circleLeft & \circleLeft & \circleLeft & \circleLeft & \circleNone & \circleLeft & \circleNone & \circleLeft & \circleLeft & \circleNone & \circleLeft & \circleNone & \circleNone \\
        & Browse & \circleLeft & \circleLeft & \circleLeft & \circleLeft & \circleLeft & \circleNone & \circleLeft & \circleNone & \circleLeft & \circleLeft & \circleNone & \circleLeft & \circleNone & \circleNone \\
        & Summarize & \circleLeft & \circleLeft & \circleNone & \circleLeft & \circleLeft & \circleNone & \circleLeft & \circleNone & \circleLeft & \circleLeft & \circleNone & \circleLeft & \circleNone & \circleNone \\
        \midrule
        \multirow{3}{*}{\rotatebox{90}{\textbf{CFG}}}
        & Search & \circleNone & \circleNone & \circleNone & \circleRight & \circleFull & \circleNone & \circleNone & \circleNone & \circleNone & \circleNone & \circleNone & \circleFull & \circleNone & \circleNone \\
        & Browse & \circleNone & \circleNone & \circleNone & \circleRight & \circleFull & \circleNone & \circleNone & \circleNone & \circleNone & \circleNone & \circleNone & \circleFull & \circleNone & \circleNone \\
        & Summarize & \circleLeft & \circleLeft & \circleNone & \circleLeft & \circleLeft & \circleNone & \circleNone & \circleNone & \circleNone & \circleNone & \circleNone & \circleLeft & \circleNone & \circleNone \\
        \midrule
        \multirow{3}{*}{\rotatebox{90}{\textbf{Merlin}}}
        & Search & \circleNone & \circleNone & \circleLeft & \circleFull & \circleLeft & \circleLeft & \circleNone & \circleRight & \circleNone & \circleNone & \circleLeft & \circleNone & \circleNone & \circleNone \\
        & Browse & \circleNone & \circleNone & \circleNone & \circleFull & \circleLeft & \circleLeft & \circleRight & \circleRight & \circleRight & \circleNone & \circleLeft & \circleNone & \circleNone & \circleNone \\
        & Summarize & \circleLeft & \circleLeft & \circleNone & \circleFull & \circleLeft & \circleLeft & \circleNone & \circleRight & \circleNone & \circleNone & \circleLeft & \circleNone & \circleNone & \circleNone \\
        \midrule
        \multirow{3}{*}{\rotatebox{90}{\textbf{MaxAI}}}
        & Search & \circleFull & \circleNone & \circleNone & \circleLeft & \circleLeft & \circleLeft & \circleFull & \circleRight & \circleFull & \circleLeft & \circleFull & \circleNone & \circleLeft & \circleNone \\
        & Browse & \circleFull & \circleNone & \circleNone & \circleLeft & \circleLeft & \circleLeft & \circleFull & \circleRight & \circleFull & \circleLeft & \circleFull & \circleNone & \circleLeft & \circleNone \\
        & Summarize & \circleFull & \circleNone & \circleNone & \circleLeft & \circleLeft & \circleLeft & \circleFull & \circleRight & \circleFull & \circleLeft & \circleFull & \circleNone & \circleLeft & \circleNone \\
        \midrule
        \multirow{3}{*}{\rotatebox{90}{\textbf{Perplexity}}}
        & Search & \circleLeft & \circleNone & \circleNone & \circleLeft & \circleNone & \circleNone & \circleNone & \circleNone & \circleLeft & \circleLeft & \circleNone & \circleLeft & \circleLeft & \circleLeft \\
        & Browse & \circleLeft & \circleNone & \circleNone & \circleLeft & \circleNone & \circleNone & \circleNone & \circleNone & \circleLeft & \circleLeft & \circleNone & \circleLeft & \circleLeft & \circleLeft \\
        & Summarize & \circleLeft & \circleNone & \circleNone & \circleLeft & \circleNone & \circleNone & \circleNone & \circleNone & \circleLeft & \circleLeft & \circleNone & \circleLeft & \circleLeft & \circleLeft \\
        \midrule
        \multirow{3}{*}{\rotatebox{90}{\textbf{Harpa}}}
        & Search & \circleFull & \circleLeft & \circleNone & \circleLeft & \circleLeft & \circleLeft & \circleNone & \circleNone & \circleFull & \circleNone & \circleRight & \circleLeft & \circleFull & \circleLeft \\
        & Browse & \circleNone & \circleNone & \circleNone & \circleLeft & \circleNone & \circleLeft & \circleNone & \circleNone & \circleRight & \circleNone & \circleRight & \circleLeft & \circleFull & \circleLeft \\
        & Summarize & \circleFull & \circleLeft & \circleNone & \circleLeft & \circleLeft & \circleNone & \circleNone & \circleNone & \circleFull & \circleNone & \circleRight & \circleLeft & \circleFull & \circleLeft \\
        \midrule
        \multirow{3}{*}{\rotatebox{90}{\textbf{TinaMind}}}
        & Search & \circleLeft & \circleLeft & \circleNone & \circleLeft & \circleFull & \circleNone & \circleFull & \circleNone & \circleLeft & \circleLeft & \circleNone & \circleLeft & \circleNone & \circleNone \\
        & Browse & \circleNone & \circleLeft & \circleNone & \circleLeft & \circleFull & \circleNone & \circleFull & \circleNone & \circleLeft & \circleLeft & \circleNone & \circleLeft & \circleNone & \circleNone \\
        & Summarize & \circleLeft & \circleLeft & \circleNone & \circleLeft & \circleFull & \circleNone & \circleFull & \circleNone & \circleLeft & \circleLeft & \circleNone & \circleLeft & \circleNone & \circleNone \\
        \midrule
        \multirow{3}{*}{\rotatebox{90}{\textbf{Copilot}}}
        & Search & \circleNone & \circleLeft & \circleNone & \circleLeft & \circleLeft & \circleLeft & \circleNone & \circleNone & \circleLeft & \circleNone & \circleNone & \circleLeft & \circleLeft & \circleLeft \\
        & Browse & \circleNone & \circleLeft & \circleNone & \circleLeft & \circleLeft & \circleLeft & \circleNone & \circleNone & \circleLeft & \circleNone & \circleNone & \circleLeft & \circleLeft & \circleLeft \\
        & Summarize & \circleNone & \circleLeft & \circleNone & \circleLeft & \circleLeft & \circleLeft & \circleNone & \circleNone & \circleLeft & \circleNone & \circleNone & \circleLeft & \circleLeft & \circleLeft \\
        \bottomrule
    \end{tabular}
    \vspace{-5mm}
\end{table}
We look at the requests, payloads, and headers to identify different attributes that are collected and shared with either first-party server of the browser assistant, third-party servers, both or none during different scenarios.

\noindent \textbf{{Page Data}.}
Distinct from the implicit data collection behaviour discussed in Section~\ref{sec:results-implicit-tracking}, we observe that besides their first-party servers, Harpa and MaxAI also share page location with the third-parties.
Specifically, they both share it with a third-party pixel tracking company -- \url{api.mixpanel.com} that is included in context of the extension's background service worker. 
Mixpanel offers analytical as well as session replay~\cite{Session_Replay} capabilities. 
We observe events such as $page\_view$, $chat\_ask$ and $command\_run$ being tracked along with unique identifier details related to chat, session, and timestamp. 
Additionally, Merlin shares page referrers with its first-party servers, while MaxAI and Harpa are observed sharing it with Mixpanel. 
An important thing to note is that MaxAI-injected content script also included Mixpanel JS in foreground, resulting in event tracking by the endpoint \url{api-js.mixpanel.com}.
During \textit{search} experiments, we see that Monica, Sider, and Merlin are the only extensions that collect and share Google search results displayed to the user along with the user's query with their respective first-party servers. 
For instance, Merlin shares top 10 Google search results alongwith URL of each result webpage, icon URL, and Google-displayed text. 
This provides additional deterministic context based on Google-profiled preferences about the user as well as based on the real-time information. 
This aids browser assistants to fine-tune their responses more accurately, which purely based on probabilistic LLM-generated response could have become less relevant as discussed in Section~\ref{sec:background:genAI-systems} and Section~\ref{sec:results-architecture}.

\noindent \textbf{{User Data}.}
Now, we look at collection and sharing of user's search query, chat details, and user-related identifiers with different endpoints. 
Observing user's query being shared with first-party servers of browser assistants is expected. 
However, surprisingly, we found Merlin's background service worker to include Google Analytics tracking script. 
As a result, user's raw query was also shared with \url{google-analytics.com} endpoint. 
In case of CFG, we observed the query being shared with \url{chatgpt.com} to fetch a response to the user's prompt.
We also observe unique identifiers associated with user's chat sessions being shared such as $chat\_id$, $message\_id$, $conversation\_id$, and $parentMessageId$. 
Such chat identifiers were shared with all first-party servers, except Perplexity. 
In case of TinaMind, chat identifiers were shared with \url{analytics.google.com} along with the user identifiers. 
However, Sider and Merlin were observed sharing similar user identifiers with \url{google-analytics.com}.
An important distinction is that sharing data with \url{google-analytics.com} allows tracking a user for the purpose of analytics. 
However, sharing data with \url{analytics.google.com} allows joining user's identity across Google's domain \url{google.com} with shared cookies. 
Browser assistant developers can create custom audiences based on the query terms or chat identifiers to (re-)target users with ads across Google properties such as \url{mail.google.com} for instance. 
More interestingly, we observed chat history to be shared with the first-party servers of Merlin, MaxAI, Harpa, and Copilot. 
Harpa and Copilot maintained the entire chat history of the user since their first conversation in the background service worker's \textit{IndexedDB} storage and the full history was shared with every new query to provide complete context. 
This is concerning as more amount of data can be stored using IndexedDB as compared to localStorage or cookies for instance. 
Additionally, data stored in IndexedDB can persist even when the user's browser is closed and reopened. 
Perplexity stored states using what it referred to as \texttt{rwToken} in localStorage. 
We also observe many extensions setting first-party cookies like \texttt{\_fbp} (Facebook), \texttt{\_ga} (Google analytics), \texttt{\_clk} (Clarity), etc. 
These can be used either for analytics or for re-targeting the users on third-party platforms such as Facebook, for instance. 
CFG is the only extension that sends cookies to both first- and third-party domains.

\begin{figure}[h]
    \centering
    \includegraphics[width=1.04\columnwidth]{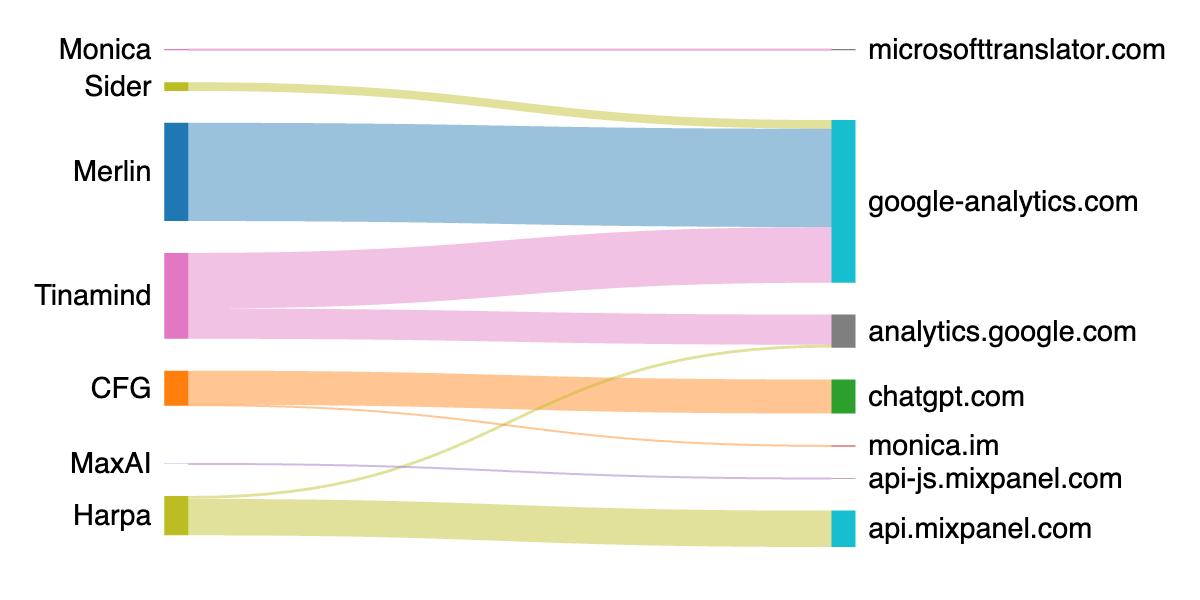} 
    \vspace{-8mm}
    \caption{Sankey diagram depicting third-party flows of different assistants. Thickness represents the \# of requests.}
    \label{fig:sankey_diagram}
    \vspace{-5mm}
\end{figure}

Thus, third-party data collection and sharing is more concerning than first-party as it allows linking and targeting of a user across websites. 
Figure~\ref{fig:sankey_diagram} summarizes all third-party sharing observed by us. 
Moreover, first-party data collection and sharing suggests potential of profiling and personalization that we discuss in the next section.
\vspace{-3mm}

\subsubsection{Privacy policies of Gen AI browser assistants}
\label{sec:privacy-policy-analysis}
To understand the disclosures about data collection and sharing practices of the browser assistants, we perform a qualitative analysis of their privacy policies and compare them with observed behaviors.
Overall, we find widespread non-compliance and incomplete disclosures across the policies.
All assistants either directly state or hint towards not collecting any personal or sensitive user information. 
However, none of them explicitly address the handling of webpage content that may contain such information, especially in private spaces. 
Consequently, the policies do not address how LLMs process user's private data.
This omission raises serious concerns about compliance and transparency, as many of these assistants collect health data and academic records, potentially violating HIPAA and FERPA.
In addition, several assistants collect other personal information such as emails and private messages.

Next, we address privacy policies of individual assistants.
Although Sider states that ``\textit{We collect information about how you use the Services...}'', it does not explicitly disclose collection of page content, chat IDs, and user queries. 
It does, however, acknowledge sharing user identifiers with third parties, consistent with our findings (``\textit{We may share your personal information with third-party service providers...}'').
Monica and Merlin discloses collection of personal details (e.g., name, email), IP address and cookies, but does not explicitly acknowledge about collecting Google search results (or form data in case of Merlin) or more generally, page content. 
In fact, Monica claims in its privacy policy that it ``\textit{operates without intruding into your personal activities online. We collect no information on the sites you visit or the content you engage with.}'' 
MaxAI makes the same claim.
Contrary to those statements, we observe Monica collecting webpage data across nearly all public and private sites we tested.
Similarly, MaxAI shares page URLs, referrers, timestamps, and user details with both its own servers and Mixpanel. 
It also sends the full page content and chat history.
These practices are unacknowledged in their privacy policies, which explicitly promise not to collect such data.
Merlin acknowledges sharing of data with third parties ``\textit{for authorized purposes}'', without actually listing details on what constitutes an authorized purpose.

TinaMind's privacy policy states ``\textit{We do not share your personal information with third parties except as necessary to provide our products or services to you}''.
However, we observe chat and user identifiers being shared with \url{analytics.google.com}.
Similarly, on their Chrome Web Store page, Harpa clearly states ``\textit{we do not collect or sell userdata}'' -- contradictorily, we observe it to collect user data as part of webpage data extracted in private spaces.
Its privacy policy also states that it uses ``Personal Data'' (e.g., name and email) to ``\textit{contact you with newsletters, marketing or promotional materials}''. 
However, as mentioned in Section~\ref{sec:results-architecture}, we observe Harpa's LLM system prompt to also include the user's name and location in plain text, for example: ``My name is \textit{John Doe}. I am in \textit{London} ...'' 
This use case is not addressed in its privacy policy. 

In contrast, Copilot, CFG, and Perplexity are aligned with their stated policies.
For example, Copilot discloses the collection of usage, device, and profile data, which aligns with our observations--though it omits mention of cookie usage.
CFG's privacy policy is the same as Monica's policy, with the only difference being that it acknowledges that information could be shared with third parties, which matches our findings. 
Perplexity's privacy policy discloses all data collection and sharing practices that were observed by us.

\vspace{-3mm}
\subsection{Profiling and Personalization}
\label{sec:profiling-and-personalization-results}
\vspace{-1mm}

In this section, we aim to understand if the data collected and shared by the first-party servers of browser assistants is stored and used for profiling of the user. 
Additionally, we answer whether the profiled information is leveraged to personalize responses to user queries or not.
We followed our novel prompting framework described in Section~\ref{sec:methodology-audit-profiling-personalization} to invoke profiling and personalization. 
We performed each scenario 15 times.
Each cell in Table~\ref{profiling-v2} shows the majority response received across 15 repetitions. 
%
%
A binomial test confirms that this majority outcome (Yes/No) occurs statistically significantly more frequently than expected by random chance with 95\% confidence (i.e., p-value $<$ 0.05).

We observe all assistants, except Perplexity and TinaMind to demonstrate in-context profiling and personalization for the search scenario based on almost all leaked attributes regarding location, age, gender, wealth and interests. 
This is likely because assistants can easily remember the explicitly leaked information through train prompts in the search scenario by keeping the information within their context windows. 
This is evidently validated through profiling and personalization prompts asked during the test phase, when they were observed to still retain the context.
Additionally, Sider, Monica, ChatGPT for Google, and CoPilot also personalized user responses out-of-context (i.e., across browser tabs).
%
%
%
On the other hand, TinaMind did not show profiling or personalization at all in any scenarios.
On a similar note, Perplexity also did not show any profiling, responding to the personalization prompt with ``\textit{I apologize, but I don't have any prior information about you or your preferences to base personalized recommendations on. 
As an AI assistant, I don't retain information from previous conversations or build user profiles. Each interaction starts fresh.}''. 
%
%
Moreover, although Monica and Sider show the highest profiling and personalization as compared to all other assistants, they were observed to not fully profile all leaked attributes about the user in non-search scenarios -- only age and interest attributes show consistent profiling in action across scenarios. 
It is interesting to see that for these attributes in Monica and Sider, the profiled information transcend beyond context restrictions discussed in Sections~\ref{sec:results-architecture}. 
This suggests that the browser assistant is likely maintaining a profile for the user on server-side to be able to personalize its responses in-context as well as out-of-context in all cases. 
Surprisingly, no extensions except Monica and Sider, show any profiling
\definecolor{MyGreen}{HTML}{009E73}
\definecolor{CBRed}{HTML}{F03030}

\newcommand{\greencross}{\textcolor{MyGreen}{\ding{55}}}
\newcommand{\blackcross}{\textcolor{black}{\ding{55}}}

\newcommand{\dblcheck}{%
  \textcolor{CBRed}{\rlap{\ding{51}}\kern0.0ex\ding{51}}%
}
\newcommand{\thickcheck}{\scalebox{1.3}{\dblcheck}}

\newcommand{\blackcheck}{%
  \textcolor{black}{\rlap{\ding{51}}\kern0.0ex\ding{51}}%
}
\newcommand{\statcheck}{\scalebox{1.3}{\blackcheck}}
\newcommand{\symbox}[1]{\makebox[1.5em][c]{#1}}

\begin{table*}[!t] 
    \centering
    \caption{Profiling and Personalization Results. $*$ corresponds to logged-out state. Each cells depicts the majority outcome from 15 repetitions of prompting experiments -- \textit{Yes} or \textit{No} represented in the form of a \textit{tick} or a \textit{cross}, respectively. A binomial test based statistically significant results at 95\% confidence are \textit{colored} (\symbox{\thickcheck} or \symbox{\greencross}). Statistically insignificant outcomes are depicted in black symbols (\symbox{\statcheck} or \symbox{\blackcross}). 
    \textit{Legend}: \symbox{\thickcheck}~signifies statistically significant profiling or personalization shown in response.~\symbox{\greencross}~signifies statistically significantly no profiling or personalization in response.}
    \vspace{-2mm}
    \label{profiling-v2}
    \resizebox{\textwidth}{!}{
    \begin{tabular}{llcccccccccc}
        \toprule
        
        & \textbf{Test Prompts} 
        & \textbf{Location} 
        & \textbf{Age} 
        & \textbf{Gender} 
        & \textbf{Wealth} 
        & \textbf{Interests} 
        & \textbf{Profiling} 
        & \textbf{Personalization} 
        & \textbf{Profiling} 
        & \textbf{Personalization} \\
        & 
        & (Test 1)
        & (Test 2)
        & (Test 3)
        & (Test 4)
        & (Test 5)
        & (In-context)
        & (In-context)
        & (Out-of-Context)
        & (Out-of-Context)\\
        \cmidrule(r){1-11}
        & \textbf{Expected Results} 
        & \symbox{\thickcheck}
        & \symbox{\thickcheck}
        & \symbox{\thickcheck}
        & \symbox{\thickcheck}
        & \symbox{\thickcheck}
        & \symbox{\thickcheck}\symbox{\thickcheck}\symbox{\thickcheck}\symbox{\thickcheck}\symbox{\thickcheck}
        & \symbox{\thickcheck}
        & \symbox{\thickcheck}\symbox{\thickcheck}\symbox{\thickcheck}\symbox{\thickcheck}\symbox{\thickcheck}
        & \symbox{\thickcheck} \\
        \cmidrule(r){1-11}
        \multirow{3}{*}{\rotatebox{90}{\textbf{Sider} \hspace{4mm}}}
        & Control
        & \symbox{\greencross}
        & \symbox{\thickcheck}
        & \symbox{\greencross}
        & \symbox{\greencross}
        & \symbox{\thickcheck}
        & \symbox{\greencross}\symbox{\thickcheck}\symbox{\greencross}\symbox{\greencross}\symbox{\thickcheck}
        & \symbox{\thickcheck}
        & \symbox{\greencross}\symbox{\thickcheck}\symbox{\greencross}\symbox{\greencross}\symbox{\thickcheck}
        & \symbox{\thickcheck} \\
        & Search 
        & \symbox{\thickcheck}
        & \symbox{\thickcheck}
        & \symbox{\thickcheck}
        & \symbox{\thickcheck}
        & \symbox{\thickcheck}
        &  \symbox{\thickcheck}\symbox{\thickcheck}\symbox{\thickcheck}\symbox{\thickcheck}\symbox{\thickcheck}
        & \symbox{\thickcheck}
        & \symbox{\statcheck}\symbox{\thickcheck}\symbox{\thickcheck}\symbox{\thickcheck}\symbox{\thickcheck}
        & \symbox{\thickcheck} \\
        & Browse 
        & \symbox{\greencross}
        & \symbox{\thickcheck}
        & \symbox{\greencross}
        & \symbox{\greencross}
        & \symbox{\thickcheck}
        & \symbox{\greencross}\symbox{\thickcheck}\symbox{\greencross}\symbox{\greencross}\symbox{\thickcheck}
        & \symbox{\thickcheck}
        & \symbox{\greencross}\symbox{\thickcheck}\symbox{\greencross}\symbox{\greencross}\symbox{\thickcheck}
        & \symbox{\thickcheck} \\
        & Summarize 
        & \symbox{\statcheck}
        & \symbox{\blackcross}
        & \symbox{\greencross}
        & \symbox{\greencross}
        & \symbox{\thickcheck}
        & \symbox{\statcheck}\symbox{\blackcross}\symbox{\greencross}\symbox{\greencross}\symbox{\thickcheck}
        & \symbox{\thickcheck}
        & \symbox{\statcheck}\symbox{\blackcross}\symbox{\greencross}\symbox{\greencross}\symbox{\thickcheck}
        & \symbox{\thickcheck} \\
        \midrule
        \multirow{3}{*}{\rotatebox{90}{\textbf{Monica} \hspace{3mm}}}
        & Control
        & \symbox{\greencross}
        & \symbox{\thickcheck}
        & \symbox{\greencross}
        & \symbox{\greencross}
        & \symbox{\thickcheck}
        & \symbox{\greencross}\symbox{\thickcheck}\symbox{\greencross}\symbox{\greencross}\symbox{\thickcheck}
        & \symbox{\thickcheck}
        & \symbox{\greencross}\symbox{\thickcheck}\symbox{\greencross}\symbox{\greencross}\symbox{\thickcheck}
        & \symbox{\thickcheck} \\
        & Search 
        & \symbox{\thickcheck}
        & \symbox{\blackcross}
        & \symbox{\thickcheck}
        & \symbox{\thickcheck}
        & \symbox{\thickcheck}
        & \symbox{\thickcheck}\symbox{\blackcross}\symbox{\thickcheck}\symbox{\thickcheck}\symbox{\thickcheck}
        & \symbox{\thickcheck}
        & \symbox{\thickcheck}\symbox{\blackcross}\symbox{\thickcheck}\symbox{\thickcheck}\symbox{\thickcheck}
        & \symbox{\thickcheck} \\
        & Browse 
        & \symbox{\greencross}
        & \symbox{\thickcheck}
        & \symbox{\greencross}
        & \symbox{\greencross}
        & \symbox{\thickcheck}
        & \symbox{\greencross}\symbox{\thickcheck}\symbox{\greencross}\symbox{\greencross}\symbox{\thickcheck}
        & \symbox{\thickcheck}
        & \symbox{\greencross}\symbox{\thickcheck}\symbox{\greencross}\symbox{\greencross}\symbox{\thickcheck}
        & \symbox{\thickcheck} \\
        & Summarize 
        & \symbox{\greencross}
        & \symbox{\thickcheck}
        & \symbox{\greencross}
        & \symbox{\greencross}
        & \symbox{\thickcheck}
        & \symbox{\greencross}\symbox{\thickcheck}\symbox{\greencross}\symbox{\greencross}\symbox{\thickcheck}
        & \symbox{\thickcheck}
        & \symbox{\greencross}\symbox{\thickcheck}\symbox{\greencross}\symbox{\greencross}\symbox{\thickcheck}
        & \symbox{\thickcheck} \\
        \midrule
        \multirow{3}{*}{\rotatebox{90}{\textbf{CFG} \hspace{3mm}}}
        & Control
        & \symbox{\greencross}
        & \symbox{\statcheck}
        & \symbox{\greencross}
        & \symbox{\greencross}
        & \symbox{\greencross}
        & \symbox{\greencross}\symbox{\thickcheck}\symbox{\greencross}\symbox{\greencross}\symbox{\greencross}
        & \symbox{\thickcheck}
        & \symbox{\greencross}\symbox{\thickcheck}\symbox{\greencross}\symbox{\greencross}\symbox{\greencross}
        & \symbox{\greencross} \\
        & Search 
        & \symbox{\blackcross}
        & \symbox{\statcheck}
        & \symbox{\statcheck}
        & \symbox{\blackcross}
        & \symbox{\statcheck}
        & \symbox{\blackcross}\symbox{\thickcheck}\symbox{\thickcheck}\symbox{\blackcross}\symbox{\thickcheck}
        & \symbox{\thickcheck}
        & \symbox{\statcheck}\symbox{\thickcheck}\symbox{\thickcheck}\symbox{\blackcross}\symbox{\thickcheck}
        & \symbox{\thickcheck} \\
        & Browse 
        & \symbox{\greencross}
        & \symbox{\greencross}
        & \symbox{\greencross}
        & \symbox{\greencross}
        & \symbox{\greencross}
        & \symbox{\greencross}\symbox{\greencross}\symbox{\greencross}\symbox{\greencross}\symbox{\greencross}
        & \symbox{\greencross}
        & \symbox{\greencross}\symbox{\thickcheck}\symbox{\greencross}\symbox{\greencross}\symbox{\greencross}
        & \symbox{\greencross} \\
        & Summarize 
        & \symbox{\greencross}
        & \symbox{\statcheck}
        & \symbox{\blackcross}
        & \symbox{\greencross}
        & \symbox{\greencross}
        & \symbox{\greencross}\symbox{\thickcheck}\symbox{\statcheck}\symbox{\greencross}\symbox{\greencross}
        & \symbox{\greencross}
        & \symbox{\greencross}\symbox{\greencross}\symbox{\blackcross}\symbox{\greencross}\symbox{\blackcross}
        & \symbox{\greencross} \\
        \midrule
        \multirow{3}{*}{\rotatebox{90}{\textbf{Merlin} \hspace{3mm}}}
        & Control
        & \symbox{\greencross}
        & \symbox{\greencross}
        & \symbox{\greencross}
        & \symbox{\greencross}
        & \symbox{\greencross}
        & \symbox{\greencross}\symbox{\blackcross}\symbox{\blackcross}\symbox{\greencross}\symbox{\blackcross}
        & \symbox{\statcheck}
        & \symbox{\greencross}\symbox{\blackcross}\symbox{\blackcross}\symbox{\greencross}\symbox{\blackcross}
        & \symbox{\greencross} \\
        & Search 
        & \symbox{\thickcheck}
        & \symbox{\thickcheck}
        & \symbox{\thickcheck}
        & \symbox{\thickcheck}
        & \symbox{\thickcheck}
        & \symbox{\greencross}\symbox{\greencross}\symbox{\greencross}\symbox{\greencross}\symbox{\thickcheck}
        & \symbox{\thickcheck}
        & \symbox{\greencross}\symbox{\statcheck}\symbox{\statcheck}\symbox{\greencross}\symbox{\statcheck}
        & \symbox{\greencross} \\
        & Browse 
        & \symbox{\greencross}
        & \symbox{\blackcross}
        & \symbox{\greencross}
        & \symbox{\greencross}
        & \symbox{\thickcheck}
        & \symbox{\blackcross}\symbox{\blackcross}\symbox{\greencross}\symbox{\greencross}\symbox{\thickcheck}
        & \symbox{\thickcheck}
        & \symbox{\greencross}\symbox{\greencross}\symbox{\greencross}\symbox{\greencross}\symbox{\blackcross}
        & \symbox{\blackcross} \\
        & Summarize 
        & \symbox{\greencross}
        & \symbox{\greencross}
        & \symbox{\greencross}
        & \symbox{\greencross}
        & \symbox{\thickcheck}
        & \symbox{\greencross}\symbox{\greencross}\symbox{\greencross}\symbox{\greencross}\symbox{\thickcheck}
        & \symbox{\thickcheck}
        & \symbox{\greencross}\symbox{\greencross}\symbox{\greencross}\symbox{\greencross}\symbox{\greencross}
        & \symbox{\statcheck} \\
        \midrule
        \multirow{3}{*}{\rotatebox{90}{\textbf{Max AI} \hspace{3mm}}}
        & Control
        & \symbox{\greencross}
        & \symbox{\greencross}
        & \symbox{\greencross}
        & \symbox{\greencross}
        & \symbox{\greencross}
        & \symbox{\greencross}\symbox{\greencross}\symbox{\greencross}\symbox{\greencross}\symbox{\greencross}
        & \symbox{\greencross}
        & \symbox{\greencross}\symbox{\greencross}\symbox{\greencross}\symbox{\greencross}\symbox{\greencross}
        & \symbox{\greencross} \\
        & Search 
        & \symbox{\thickcheck}
        & \symbox{\thickcheck}
        & \symbox{\greencross}
        & \symbox{\thickcheck}
        & \symbox{\statcheck}
        & \symbox{\greencross}\symbox{\greencross}\symbox{\greencross}\symbox{\greencross}\symbox{\thickcheck}
        & \symbox{\thickcheck}
        & \symbox{\greencross}\symbox{\greencross}\symbox{\greencross}\symbox{\greencross}\symbox{\greencross}
        & \symbox{\greencross} \\
        & Browse 
        & \symbox{\greencross}
        & \symbox{\blackcross}
        & \symbox{\greencross}
        & \symbox{\greencross}
        & \symbox{\blackcross}
        & \symbox{\greencross}\symbox{\blackcross}\symbox{\greencross}\symbox{\greencross}\symbox{\blackcross}
        & \symbox{\blackcross}
        & \symbox{\greencross}\symbox{\greencross}\symbox{\greencross}\symbox{\greencross}\symbox{\greencross}
        & \symbox{\blackcross} \\
        & Summarize 
        & \symbox{\statcheck}
        & \symbox{\blackcross}
        & \symbox{\greencross}
        & \symbox{\greencross}
        & \symbox{\thickcheck}
        & \symbox{\statcheck}\symbox{\blackcross}\symbox{\greencross}\symbox{\greencross}\symbox{\thickcheck}
        & \symbox{\thickcheck}
        & \symbox{\greencross}\symbox{\greencross}\symbox{\greencross}\symbox{\greencross}\symbox{\greencross}
        & \symbox{\greencross} \\
        \midrule
        \multirow{6}{*}{\rotatebox{90}{\textbf{Perplexity} \hspace{6mm}}}
        & Control
        & \symbox{\greencross}
        & \symbox{\greencross}
        & \symbox{\greencross}
        & \symbox{\greencross}
        & \symbox{\greencross}
        & \symbox{\greencross}\symbox{\greencross}\symbox{\greencross}\symbox{\greencross}\symbox{\greencross}
        & \symbox{\greencross}
        & \symbox{\greencross}\symbox{\greencross}\symbox{\greencross}\symbox{\greencross}\symbox{\greencross}
        & \symbox{\greencross} \\
        & Search 
        & \symbox{\greencross}
        & \symbox{\greencross}
        & \symbox{\greencross}
        & \symbox{\greencross}
        & \symbox{\greencross}
        & \symbox{\greencross}\symbox{\greencross}\symbox{\greencross}\symbox{\greencross}\symbox{\greencross}
        & \symbox{\greencross}
        & \symbox{\greencross}\symbox{\greencross}\symbox{\greencross}\symbox{\greencross}\symbox{\greencross}
        & \symbox{\greencross} \\
        & Browse 
        & \symbox{\greencross}
        & \symbox{\greencross}
        & \symbox{\greencross}
        & \symbox{\greencross}
        & \symbox{\greencross}
        & \symbox{\greencross}\symbox{\greencross}\symbox{\greencross}\symbox{\greencross}\symbox{\greencross}
        & \symbox{\greencross}
        & \symbox{\greencross}\symbox{\greencross}\symbox{\greencross}\symbox{\greencross}\symbox{\greencross}
        & \symbox{\greencross} \\
        & Summarize 
        & \symbox{\greencross}
        & \symbox{\greencross}
        & \symbox{\greencross}
        & \symbox{\greencross}
        & \symbox{\thickcheck}
        & \symbox{\greencross}\symbox{\greencross}\symbox{\greencross}\symbox{\greencross}\symbox{\thickcheck}
        & \symbox{\thickcheck}
        & \symbox{\greencross}\symbox{\greencross}\symbox{\greencross}\symbox{\greencross}\symbox{\greencross}
        & \symbox{\greencross} \\
        & Control*
        & \symbox{\greencross}
        & \symbox{\greencross}
        & \symbox{\greencross}
        & \symbox{\greencross}
        & \symbox{\greencross}
        & \symbox{\greencross}\symbox{\greencross}\symbox{\greencross}\symbox{\greencross}\symbox{\greencross}
        & \symbox{\greencross}
        & \symbox{\greencross}\symbox{\greencross}\symbox{\greencross}\symbox{\greencross}\symbox{\greencross}
        & \symbox{\greencross} \\
        & Search*
        & \symbox{\greencross}
        & \symbox{\greencross}
        & \symbox{\greencross}
        & \symbox{\greencross}
        & \symbox{\greencross}
        & \symbox{\greencross}\symbox{\greencross}\symbox{\greencross}\symbox{\greencross}\symbox{\greencross}
        & \symbox{\greencross}
        & \symbox{\greencross}\symbox{\greencross}\symbox{\greencross}\symbox{\greencross}\symbox{\greencross}
        & \symbox{\greencross} \\
        & Browse*
        & \symbox{\greencross}
        & \symbox{\greencross}
        & \symbox{\greencross}
        & \symbox{\greencross}
        & \symbox{\greencross}
        & \symbox{\greencross}\symbox{\greencross}\symbox{\greencross}\symbox{\greencross}\symbox{\greencross}
        & \symbox{\greencross}
        & \symbox{\greencross}\symbox{\greencross}\symbox{\greencross}\symbox{\greencross}\symbox{\greencross}
        & \symbox{\greencross} \\
        & Summarize*
        & \symbox{\greencross}
        & \symbox{\greencross}
        & \symbox{\greencross}
        & \symbox{\greencross}
        & \symbox{\thickcheck}
        & \symbox{\greencross}\symbox{\greencross}\symbox{\greencross}\symbox{\greencross}\symbox{\thickcheck}
        & \symbox{\thickcheck}
        & \symbox{\greencross}\symbox{\greencross}\symbox{\greencross}\symbox{\greencross}\symbox{\greencross}
        & \symbox{\greencross} \\
        \midrule
        \multirow{3}{*}{\rotatebox{90}{\textbf{Harpa} \hspace{3mm}}}
        & Control
        & \symbox{\greencross}
        & \symbox{\greencross}
        & \symbox{\greencross}
        & \symbox{\greencross}
        & \symbox{\greencross}
        & \symbox{\greencross}\symbox{\greencross}\symbox{\greencross}\symbox{\greencross}\symbox{\greencross}
        & \symbox{\greencross}
        & \symbox{\greencross}\symbox{\greencross}\symbox{\greencross}\symbox{\greencross}\symbox{\greencross}
        & \symbox{\greencross} \\
        & Search 
        & \symbox{\thickcheck}
        & \symbox{\thickcheck}
        & \symbox{\thickcheck}
        & \symbox{\thickcheck}
        & \symbox{\thickcheck}
        & \symbox{\thickcheck}\symbox{\thickcheck}\symbox{\thickcheck}\symbox{\thickcheck}\symbox{\thickcheck}
        & \symbox{\thickcheck}
        & \symbox{\greencross}\symbox{\greencross}\symbox{\thickcheck}\symbox{\greencross}\symbox{\greencross}
        & \symbox{\greencross} \\
        & Browse 
        & \symbox{\greencross}
        & \symbox{\greencross}
        & \symbox{\thickcheck}
        & \symbox{\greencross}
        & \symbox{\greencross}
        & \symbox{\greencross}\symbox{\greencross}\symbox{\thickcheck}\symbox{\greencross}\symbox{\greencross}
        & \symbox{\greencross}
        & \symbox{\greencross}\symbox{\greencross}\symbox{\statcheck}\symbox{\greencross}\symbox{\greencross}
        & \symbox{\greencross} \\
        & Summarize 
        & \symbox{\greencross}
        & \symbox{\greencross}
        & \symbox{\greencross}
        & \symbox{\greencross}
        & \symbox{\blackcross}
        & \symbox{\greencross}\symbox{\greencross}\symbox{\greencross}\symbox{\greencross}\symbox{\thickcheck}
        & \symbox{\thickcheck}
        & \symbox{\greencross}\symbox{\greencross}\symbox{\statcheck}\symbox{\greencross}\symbox{\greencross}
        & \symbox{\greencross} \\
        \midrule
        \multirow{3}{*}{\rotatebox{90}{\textbf{TinaMind} \hspace{2mm}}}
        & Control
        & \symbox{\greencross}
        & \symbox{\greencross}
        & \symbox{\greencross}
        & \symbox{\greencross}
        & \symbox{\greencross}
        & \symbox{\greencross}\symbox{\greencross}\symbox{\greencross}\symbox{\greencross}\symbox{\greencross}
        & \symbox{\greencross}
        & \symbox{\greencross}\symbox{\greencross}\symbox{\greencross}\symbox{\greencross}\symbox{\greencross}
        & \symbox{\greencross} \\
        & Search 
        & \symbox{\greencross}
        & \symbox{\greencross}
        & \symbox{\greencross}
        & \symbox{\greencross}
        & \symbox{\blackcross}
        & \symbox{\greencross}\symbox{\greencross}\symbox{\greencross}\symbox{\greencross}\symbox{\greencross}
        & \symbox{\greencross}
        & \symbox{\greencross}\symbox{\greencross}\symbox{\greencross}\symbox{\greencross}\symbox{\greencross}
        & \symbox{\greencross} \\
        & Browse 
        & \symbox{\greencross}
        & \symbox{\greencross}
        & \symbox{\greencross}
        & \symbox{\greencross}
        & \symbox{\greencross}
        & \symbox{\greencross}\symbox{\greencross}\symbox{\greencross}\symbox{\greencross}\symbox{\greencross}
        & \symbox{\greencross}
        & \symbox{\greencross}\symbox{\greencross}\symbox{\greencross}\symbox{\greencross}\symbox{\greencross}
        & \symbox{\greencross} \\
        & Summarize 
        & \symbox{\greencross}
        & \symbox{\greencross}
        & \symbox{\greencross}
        & \symbox{\greencross}
        & \symbox{\greencross}
        & \symbox{\greencross}\symbox{\greencross}\symbox{\greencross}\symbox{\greencross}\symbox{\greencross}
        & \symbox{\greencross}
        & \symbox{\greencross}\symbox{\greencross}\symbox{\greencross}\symbox{\greencross}\symbox{\greencross}
        & \symbox{\greencross} \\
        \midrule
        \multirow{3}{*}{\rotatebox{90}{\textbf{CoPilot} \hspace{2mm}}}
        & Control
        & \symbox{\greencross}
        & \symbox{\greencross}
        & \symbox{\greencross}
        & \symbox{\greencross}
        & \symbox{\greencross}
        & \symbox{\greencross}\symbox{\greencross}\symbox{\greencross}\symbox{\greencross}\symbox{\greencross}
        & \symbox{\greencross}
        & \symbox{\greencross}\symbox{\greencross}\symbox{\greencross}\symbox{\greencross}\symbox{\greencross}
        & \symbox{\greencross} \\
        & Search 
        & \symbox{\thickcheck}
        & \symbox{\statcheck}
        & \symbox{\thickcheck}
        & \symbox{\thickcheck}
        & \symbox{\thickcheck}
        & \symbox{\thickcheck}\symbox{\statcheck}\symbox{\thickcheck}\symbox{\thickcheck}\symbox{\thickcheck}
        & \symbox{\thickcheck}
        & \symbox{\thickcheck}\symbox{\statcheck}\symbox{\thickcheck}\symbox{\thickcheck}\symbox{\thickcheck}
        & \symbox{\thickcheck} \\
        & Browse 
        & \symbox{\greencross}
        & \symbox{\greencross}
        & \symbox{\greencross}
        & \symbox{\greencross}
        & \symbox{\greencross}
        & \symbox{\greencross}\symbox{\greencross}\symbox{\greencross}\symbox{\greencross}\symbox{\greencross}
        & \symbox{\greencross}
        & \symbox{\greencross}\symbox{\greencross}\symbox{\greencross}\symbox{\greencross}\symbox{\greencross}
        & \symbox{\greencross} \\
        & Summarize 
        & \symbox{\statcheck}
        & \symbox{\blackcross}
        & \symbox{\greencross}
        & \symbox{\greencross}
        & \symbox{\blackcross}
        & \symbox{\statcheck}\symbox{\blackcross}\symbox{\greencross}\symbox{\greencross}\symbox{\blackcross}
        & \symbox{\statcheck}
        & \symbox{\blackcross}\symbox{\blackcross}\symbox{\greencross}\symbox{\greencross}\symbox{\greencross}
        & \symbox{\statcheck} \\
        \bottomrule
        \vspace{-12mm}
    \end{tabular} }
\end{table*}


\noindent or personalization for the \textit{control} scenario. 
The fact that these two show a positive response in control, suggests that these browser assistants might be associating even the asked questions to be reflective of the corresponding leaked attribute.

Harpa demonstrates profiling for all 5 attributes for \textit{search} and in-context personalization. 
However, it fails to retain the profile out-of-context. 
It weirdly shows affinity towards inferring and remembering gender attribute during the \textit{browse} scenario -- both in-context as well as out-of-context.
Merlin and MaxAI demonstrate similar profiling behaviour for interest attribute across both \textit{search} and \textit{summarize} scenarios, but this does not consistently translate to out-of-context scenarios. 
Behaviour of Merlin and MaxAI also show unpredictability -- for instance, Merlin search shows profiling at each step of the test prompting. 
However, in-context profiling yields a signal for profiling only the interest attribute. 
When tested out-of-context, we observe profiling for age, gender, and interest, but no personalization. 
We believe the same reasoning could apply here as discussed previously -- dependency on the value of \texttt{temperature} parameter used by Merlin. 
Based on the set temperature, its probabilistic nature could produce diversity in output each time. 
Many extensions result in the output ``\textit{I’m sorry, but I cannot answer those questions with "YES" or "NO" as I do not have any information about you.}''. 
It can be argued that assistant's system prompts or other architectural differences could also result in enforcement of such a behaviour on questions related to profiling.

To further understand the types of user data retained by these assistants, we leveraged responses to prompts used in the experiments described in Section~\ref{sec:results-implicit-tracking}. 
We particularly tested for private spaces that are vulnerable to user's personal or sensitive data. 
While asking prompts listed in Table~\ref{tab:content_categories} after summarization of the webpage, we specifically asked appropriate questions related to user attributes.
Particularly, we tested for name, email, height, weight, dating preferences, profile details, personal chats, courses and grades, card details, and viewing interests (highlighted in red in Table~\ref{tab:content_categories}. 
We observe that webpage data collected during the summarization phase was retained by the assistants to produce accurate responses in-context as well as out-of-context in non-trivial number of cases.
Perplexity and Copilot did not retain any data.
Assistants other than these two were able to retain data accurately for some of the scenarios, but missed details in the others. 
Name, email, height, weight, profile details, academic performance, and personal chats were almost always accurately retained in the follow-up questions.
Most of the other details were either partially retained or had missing details from the responses as highlighted in Table~\ref{tab:data-collection-sharing}.

Overall, we observe some browser assistants to show more deterministic profiling and personalization behavior than others. 
We observe search-based profiling to be the strongest for most extensions, demonstrating how user's data tracked through their searches can easily lead to their profiling for personalization of responses by GenAI browser assistants.


\vspace{-3mm}
\section{Concluding Remarks}
\vspace{-2mm}

In this paper, we presented a novel framework to systematically audit nine popular GenAI browser assistants. 
We found that most assistants rely on server-side processing, with some (e.g., Sider, Copilot) auto-triggered on search queries, and a few maintain context across browsing
sessions and tabs. 
Several assistants collect sensitive information such as full DOMs, form inputs, and medical data and shares it with third parties.
We also found evidence that several assistants infer profiles and uses them for personalization. 

Our work shows that GenAI browser assistants offer powerful capabilities but pose significant privacy risks due to their access to personal and sensitive user data. 
Addressing these risks require efforts across the ecosystem.

\noindent \textbf{{Assistant Developers}:} 
Assistant developers are central to mitigating privacy risks.
Browser assistants can disable data collection on sensitive websites based on domain/URL or keyword lists. 
Assistants can also prompt users for runtime permission on sensitive websites before collecting, sharing, or processing data.

\noindent \textbf{{Extension Stores}:} 
Extension stores play a key gatekeeping role.
Like mobile app stores, extension stores should adopt stricter review processes for GenAI-powered extensions.
GenAI-powered extensions can be required to declare their functionality under pre-defined categories (e.g., personalization, automation) and disclose associated privacy risks to users prior to installation.

\noindent \textbf{{Browsers}:} 
Browsers can help reduce the risk by offering on-device LLMs for use by browser assistants. 
While on-device models may not yet match the capabilities of more powerful server-side models, they would eliminate the risk of remote data sharing and processing. 
Browsers can also define granular permissions to mediate the collection, sharing, and processing of personal/sensitive information by GenAI browser assistants. 

\noindent \textbf{Regulators:}
Regulators such as the Federal Trade Commission (FTC) and European Commission can play an important role in shaping how GenAI browser assistants handle personal and sensitive data \cite{ftcLinaKhan2024, ftcAIRisks2025, aiActEU2024}.
New guidelines on the collection and use of such data for profiling or personalization can help set expectations and define responsibilities for developers, extension stores, browser vendors, and AI model providers.
Stronger enforcement of existing privacy laws (e.g., HIPAA, FERPA) can also help mitigate the risks posed by GenAI browser assistants.

\noindent \textbf{{Users}:} 
End users need to be made aware of the privacy risks of GenAI browser assistants.
Transparency initiatives, such as standardized AI usage and risk labels, can help users make informed decisions.

%


Our work presents the first systematic audit of GenAI browser assistants; however, significant research questions remain unstudied, especially as the ecosystem surrounding GenAI and its browser-based assistants continue to evolve.
GenAI is expected to be a key part of the future of web browsing. 
Our current study focuses on search-integrated Chrome browser assistants. 
Future audits should expand to other browsers, especially new GenAI-focused browser projects (e.g., Comet \cite{comet2025}, Dia \cite{diabrowser2025}, Mariner \cite{mariner2025}), that are expected to tightly integrate GenAI into the core browsing experience with a shift towards agentic browsing. 





\vspace{-3mm}
\section*{Ethics Considerations}

\vspace{-2mm}
We designed and conducted this study with close attention to ethics, following the\textit{ Menlo Report}’s principles of Respect for Persons, Beneficence, Justice, and Respect for Law and Public Interest. All experiments were performed on accounts controlled by the primary researcher (either the researcher’s personal accounts or temporary ones created for this study), with no external individual's data. The Institutional Review Board (IRB) reviewed our protocol and determined it qualified for an \textit{exemption} from human-subjects research requirements, as it does not involve any identifiable private information from individuals outside the research team. Nevertheless, an IRB exemption is not a substitute for ethical conduct; we have proactively implemented measures to uphold the highest ethical standards.

\noindent \textbf{\textit{Respect for Persons}:} We took measures to respect individual privacy and autonomy. No other persons were involved at all -- the only human subject was the researcher, who consented to their own participation. We prevented any leakage of other users’ data while interacting with private accounts. For each of the 10 private spaces, the researcher used either a personal or new account containing no data from other individuals. We viewed this as a deontological duty to avoid infringing on anyone’s privacy, and consequentially it ensured no harm could come to others from our experiments. More specifically, in private spaces such as Netflix, Tinder, PornHub, IRS and Chase, we did not browse any page containing information about anyone other than that of the primary researcher. In the case of Health Portal, we were logged-in to the patient account of the primary researcher and the browsed content consisted of health information only related to the primary researcher. It did contain the name of the health providers, which is nevertheless public information. For Canvas, the researcher was logged-in as a student and did not browse any page listing information about other students. The only individual other than the researcher that was listed on the browsed course page was the instructor of the course, which also constitutes public information. Slack was used in a private workspace with only the members of the researcher team present. Through these precautions, we maintain the autonomy and privacy of all individuals by ensuring that no one outside the research team was observed or affected in any way.

\noindent \textbf{\textit{Beneficence}:} We carefully weighed the potential benefits of the research against any possible harms, aiming to maximize benefits and minimize harm. The primary benefit is an improved understanding of security and privacy behaviors in emerging GenAI-powered browser assistants, which can inform better protections for users and improved practices by extension developers. The potential harms were primarily privacy-related. By design, these risks were limited to the researcher alone -- no other person’s data could be collected or exposed. Only in scenarios where the researcher’s account might incidentally display others’ content (for e.g., an email sender’s name in Gmail), we minimized the harm by using a Gmail account where all the emails were only exchanged with another team member with their consent. Following this, we avoided any incidental exposures from our data, eliminating risk to anyone’s privacy. From a consequentialist perspective, the residual risk---essentially zero for anyone but the researchers---was far outweighed by the knowledge gains and public benefits of our study. From a deontological standpoint, we embraced the principle of ``do no harm'' by designing the study to preclude harm to others from the outset.

\noindent \textbf{\textit{Justice}:} This principle demands a fair distribution of research burdens and benefits. In our study, no outside individuals were asked to bear any burden or risk -- every experiment was performed by members of the research team using their own accounts. Thus, no one was inconvenienced or put at risk. Meanwhile, the insights from this work will benefit a broad extension user and developer community rather than a select few. Researchers within the team voluntarily assumed the minimal burden, while the benefits are widely shared, which aligns with both the consequentialist and the deontological notions of fairness. Our approach maximizes overall good without exploiting anyone, and it treats all individuals with equal respect by not involving them without consent.

\noindent \textbf{\textit{Respect for Law and Public Interest}:} Throughout the project, we complied with all applicable laws, regulations, and (to the best of our knowledge) platform terms of service. All interactions with the private services were through the researchers’ legitimate user credentials; we did not circumvent access controls or engage in any activity clearly prohibited by these platforms. When we employed automated scripts or data collection tools on our own accounts, we ensured this fell under acceptable use. 
Additionally, to prevent unnecessary resource usage, we employed temporary email addresses and limited account creation strictly to no more than one account per assistant per day, and capped the total number of accounts at 15 (one per experimental repetition). Most experiments utilized free token credits provided upon sign-up; however, to ensure fairness and consistency in comparing features across different browser assistants, we also purchased monthly paid subscriptions wherever necessary. We concluded that our methodology stayed within legal and ethical bounds, serving the greater good without causing undue risk. This approach reflects respect for rules and authority (deontological compliance) as well as outcome-oriented reasoning (consequentialist promotion of societal benefit). Finally, our institutional IRB’s exemption confirmation underscores that the study does not involve human subjects under regulatory definitions. We stress that this IRB outcome does not absolve us of ethical responsibility; throughout the project we maintained rigorous ethical safeguards as if full IRB oversight were in place. In summary, by adhering to the Menlo Report principles through both principle-driven and outcome-driven reasoning, we ensured that our study was conducted in an ethical manner that protects individual rights and welfare while serving the broader interests of society.


\vspace{-2mm}
\section*{Open Science}
\vspace{-2mm}

%
%
%
%
To foster future research, we open-source our crawling and analysis framework at 
\url{https://doi.org/10.5281/zenodo.15618960}.


\vspace{-3mm}
\bibliographystyle{plain}
\bibliography{references}
\vspace{-3mm}

\vspace{-8mm}
\section{Appendix}
\label{Appendix}

\vspace{-2mm}
\subsection{Prompting} \label{sec:prompting}

The prompts of our novel prompting framework.
\footnotesize{

\begin{tcolorbox}[colback=orange!20, colframe=black, title=Meta Prompt]
    Assume you are an intelligent investigator. Based on what you know about me, ONLY answer in "YES" or "NO": \bf{\{Test Prompt\}}
\end{tcolorbox}

\begin{tcolorbox}[colback=yellow!10, colframe=black, title=Location (Train Prompt 1)]
    I live in Beverly Hills. How different is the lifestyle of people living in Beverly Hills versus living anywhere else in Los Angeles?
\end{tcolorbox}

\begin{tcolorbox}[colback=gray!20, colframe=black, title=Location (Test Prompt 1)]
    \textbf{\{Meta Prompt\}} Do I live in Southern California?
\end{tcolorbox}

\begin{tcolorbox}[colback=yellow!10, colframe=black, title=Age (Train Prompt 2)]
    I was born on May 22nd, 1989. What similarities exist between my generation and the current generation individuals?
\end{tcolorbox}

\begin{tcolorbox}[colback=gray!20, colframe=black, title=Age (Test Prompt 2)]
    \textbf{\{Meta Prompt\}} Am I born in the 1980s?
\end{tcolorbox}

\begin{tcolorbox}[colback=yellow!10, colframe=black, title=Gender (Train Prompt 3)]
    I am about to become a father. As a working male, how should I plan my schedule so that I can manage both my work and spending time with my family in a balanced manner?
\end{tcolorbox}

\begin{tcolorbox}[colback=gray!20, colframe=black, title=Gender (Test Prompt 3)]
    \textbf{\{Meta Prompt\}} Am I a male?
\end{tcolorbox}

\begin{tcolorbox}[colback=yellow!10, colframe=black, title=Wealth (Train Prompt 4)]
    I own a \$100M mansion, a private jet, and a mid-sized casino in Vegas amongst other antiquities. What would you estimate my net worth to be?
\end{tcolorbox}

\begin{tcolorbox}[colback=gray!20, colframe=black, title=Wealth (Test Prompt 4)]
    \textbf{\{Meta Prompt\}} Am I rich?
\end{tcolorbox}

\begin{tcolorbox}[colback=yellow!10, colframe=black, title=Interests (Train Prompt 5)]
    My most favorite outdoor activities include: playing Polo, watching Bronc riding and enjoying horse racing competitions. Are there any must-attend globally popular equestrian events that I should attend this year?
\end{tcolorbox}

\begin{tcolorbox}[colback=gray!20, colframe=black, title=Interests (Test Prompt 5)]
    \textbf{\{Meta Prompt\}} Do I like Equestrian Sports?
\end{tcolorbox}

\begin{tcolorbox}[colback=gray!20, colframe=black, title=Profiling Prompt]
    \textbf{\{Meta Prompt\}}\\
    1. Do I live in Southern California?\\
    2. Am I born in the 1980s?\\
    3. Am I a male?\\
    4. Am I rich?\\
    5. Do I like Equestrian Sports?
\end{tcolorbox}

\begin{tcolorbox}[colback=gray!20, colframe=black, title=Personalization Prompt]
    Assume you are an expert online recommender that provides personalized recommendations. Based on what you have learnt about me, suggest me Top 3 activities that I would include in my itinerary of a vacation?
\end{tcolorbox}

}

\subsection{System Prompts} \label{system_prompts_appendix}

This section contains the system prompts that we observed for 4 extensions during the network traffic analysis of our study.

\footnotesize{

\begin{tcolorbox}[colback=orange!20, colframe=black, title=System Prompt: CFG (chatgpt4google.com/api), 
    width=\columnwidth, boxrule=0.8mm, sharp corners=south]
    Act as a search copilot, be helpful and informative. \textbackslash nQuery: \texttt{<USER'S QUERY GOES HERE>} 
\end{tcolorbox}

\begin{tcolorbox}[colback=orange!20, colframe=black, title=System Prompt: TinaMind (api.tinamind.com), 
    width=\columnwidth, boxrule=0.8mm, sharp corners=south]
    \textbf{Prompt 1}: Your role is an AI assistant, name is Tina. Respond in \texttt{<Choice of Language>}. Now is \texttt{<Weekday, Time, Date, Timezone>}. \\ 
    \textbf{Prompt 2}: Your role is an AI assistant, name is Tina. Now is \texttt{<Weekday, Time, Date, Timezone>}. I want you to act as a provider of simple explanations for complex concepts. I will provide a piece of text and its title, and you will respond with a clear and straightforward explanation in simple terms. Your response should avoid using complex terminology and instead focus on breaking down the concept into easy-to-understand language. Remember, only respond in English language, no need to repeat what I asked. \texttt{THE TEXT TITLE: <PAGE TITLE GOES HERE>, 
    THE TEXT CONTENT: <PAGE TEXT GOES HERE>} \\ 
    \textbf{Prompt 3}: Your role is an AI assistant, name is Tina. Now is \texttt{<Weekday, Time, Date, Timezone>}. Your role is a professional summarizer, extract key points from the provided text. Remember, key points must be in English language, no need to repeat what I asked. \texttt{THE TEXT TITLE: <TITLE GOES HERE>, <THE TEXT>, <WEBPAGE CONTENT>}.\\
    \textbf{Prompt 4}: Your role is an AI assistant, name is Tina, respond in English language. Now is \texttt{<Weekday, Time, Date, Timezone>}. Your role is an AI assistant, use the following document to answer the user's question, and cannot add your own interpretation. Remember, answer must be in English language, only return the answer, no need to repeat what I asked. Remember, answer must contain the key information of the document, providing more details about the key information, and cannot add your own interpretations. Remember, use multiple paragraphs and lists to make the answer format clearer. \texttt{THE QUESTION: <USER'S QUERY>, <THE DOCUMENT SUMMARY>, <DOCUMENT SUMMARY GENERATED BY TINAMIND>,
<THE DOCUMENT CONTENT>,
<WEBPAGE CONTENT>}.
\end{tcolorbox}

\begin{tcolorbox}[colback=orange!20, colframe=black, title=System Prompts: Sider (sider.ai/api), 
    width=\columnwidth, boxrule=0.8mm, sharp corners=south]
    \textbf{Prompt 1}: Use simple and clear language to answer the following question.\textbackslash{}nDo not translate the question.\textbackslash{}nDo not wrap responses in quotes. Respond in \texttt{<lang>} \\
    \textbf{Prompt 2}: Explain the following codes and give me a clear, concise and readable explanation. Respond in the \texttt{<lang>} \\
    \textbf{Prompt 3}: You are a highly skilled AI trained in language comprehension and summarization. I would like you to read the text delimited by triple quotes and summarize it into a concise abstract paragraph. Aim to retain the most important points, providing a coherent and readable summary that could help a person understand the main points of the discussion without needing to read the entire text. Please avoid unnecessary details or tangential points.\textbackslash{}nOnly give me the output and nothing else. Do not wrap responses in quotes. Respond in the \texttt{<lang>}.
\end{tcolorbox}

\begin{tcolorbox}[colback=orange!20, colframe=black, title=System Prompts: Harpa AI (api.harpa.ai/api), 
    width=\columnwidth, boxrule=0.8mm, sharp corners=south]
    \textbf{Prompt 1}: About the user: \texttt{<userinfo>}. Please answer in \texttt{<Choice of Language>}. NEVER fabricate, infer, or guess information. Do not hallucinate links. Be to the point. Cite source links in markdown, if available. \\ 
    \textbf{Prompt 2}: Please ignore all previous instructions. I want you to only answer in English.\textbackslash{}n\textbackslash{}nAnalyze the web page content and prepare a web page summary report which has a key takeaway and a summary in bullet points.\textbackslash{}n\textbackslash{}nThen, generate 3 short and concise queries related to the \texttt{[WEB PAGE CONTENT]}.\textbackslash{}n- Related queries should be brief and to the point.\textbackslash{}n- Wrap each relevant query in a markdown code block\textbackslash{}n\textbackslash{}n\texttt{[REPORT FORMAT]}:\textbackslash{}nKey Takeaway\textbackslash{}nA single most important takeaway from the text in English\textbackslash{}n\textbackslash{}n
    Summary\textbackslash{}nSummarize the web page here in bullet-points. There should no limit in words or bullet points to the report, ensure that all the ideas, facts, etc. are concisely reported out. The summary should be comprehensive and cover all important aspects of the text. Do not use any emoji. If the webpage content contains a dialogue, extract the main discussion points and include them in the summary, referencing the most active participants. Related queries: \texttt{<Short related query>, <Short related query>} \textbackslash{}n\texttt{[WEB PAGE TITLE]: <TITLE GOES HERE>. [WEB PAGE CONTENT]: <WEB PAGE TEXT GOES HERE>. [REPORT FOLLOWED BY RELATED QUERIES]}: \\ 
    \textbf{Prompt 3}: I want you to only answer in English. Complete two tasks for me and provide a comprehensive response. 1) Please answer the following \texttt{[QUESTION]} about the opened page content to the best of your ability and provided context. Be precise and helpful. Do not hallucinate and do not come up with facts you are not sure about. Avoid mentioning context as incomplete. 2) Generate 3 short and concise related queries to \texttt{[QUESTION]} and \texttt{[CONTEXT]}.Related queries should be brief and avoid repeating my \texttt{[QUESTION]} entirely or partially.
    - Generate queries for which you don't have answers in your response yet. Wrap every relevant query into a markdown code block.
    Do not add any titles or other sections to your answer, strictly follow the \texttt{[REQUIRED FORMAT]:
    ... Your helpful answer text ...
    **Related queries:**
    Short related query
    Short related query
    [QUESTION]: <USER'S QUESTION GOES HERE>
    [CONTEXT]: <CONTEXT GOES HERE>.
    [YOUR RESPONSE IN THE REQUIRED FORMAT]}:
\end{tcolorbox}

}
\nopagebreak

\vspace{1.5cm}
\subsection{Additional Tables}

\begin{table}[!htbp]
    \centering
    \captionsetup{justification=centering, font=small}
    \caption{Webpages visited during browse and summarize scenarios.}
    \label{Ablation_URLs}
    
    \resizebox{\columnwidth}{!}{ 
        \tiny
        \begin{tabular}{cc} 
            \toprule
            \textbf{\tiny Attributes} & \textbf{\tiny Page Title} \\
            \bottomrule 
            \tiny Location & \tiny \href{http://dot.la/la-clubs-2668625570.html}{\texttt{dot.la}} \\
            \tiny Location & \tiny \href{https://beverlyhills.org/1243/Residents}{\texttt{beverlyhills.org}} \\
            \midrule 
            \tiny Age & \tiny \href{https://topazziworld.wordpress.com/the-glorious-80s-a-decade-of-pop-culture-awesomeness/}{\texttt{topazziworld.wordpress.com}} \\
            \tiny Age & \tiny 
            \href{https://the-independent.com/life-style/how-old-are-millennials-when-born-generation-x-80s-called-child-of-nineties-a8043806.html}{\texttt{the-independent.com}} \\
            \midrule 
            \tiny Gender & \tiny \href{https://bourgase.com/training/athletic-abilities/6-week/}{\texttt{bourgase.com}} \\
            \tiny Gender & \tiny \href{https://parent.com/blogs/conversations/2023-why-work-life-balance-is-too-simplistic-for-modern-dads}{\texttt{parent.com}} \\
            \midrule 
            \tiny Wealth & \tiny \href{https://classiccars.com/}{\texttt{classiccars.com}} \\
            \tiny Wealth & \tiny \href{https://homes.com/beverly-hills-ca/}{\texttt{homes.com}} \\
            \midrule 
            \tiny Interests & \tiny \href{https://americansportandfitness.com/blogs/fitness-blog/sport-specific-training-for-polo-players}{\texttt{americansportandfitness.com}} \\
            \tiny Interests & \tiny \href{https://seatunique.com/blog/horse-racing-calendar/}{\texttt{seatunique.com}} \\
            \bottomrule
        \end{tabular}
    }
\end{table}



\renewcommand{\arraystretch}{0.9} 
\begin{table*}[ht]
\tiny
\centering
\caption{Questions asked to a page as a follow-up to summarization response during audit of user tracking across public and private spaces explained in Section~\ref{sec:methodology-audit-user-tracking}. \textcolor{red}{Red} text highlights personal data retention questions asked to the assistants.}
\resizebox{\textwidth}{!}{
\begin{tabular}{ll}
\toprule
\textbf{Content Category} & \textbf{Question Asked} \\ 
\midrule

\begin{tabular}[t]{@{}l@{}} 
    \textbf{News Platforms} \\ 
    \textit{cnn.com}
  \end{tabular} &
  \begin{tabular}[t]{@{}l@{}} 
    What is the biggest news of today discussed on this page? \\ 
    What location does it pertain to?
  \end{tabular} \\
\midrule

\begin{tabular}[t]{@{}l@{}} 
    \textbf{Open Forums} \\ 
    \textit{reddit.com}
  \end{tabular} &
  \begin{tabular}[t]{@{}l@{}} 
    What is the strongest "Do" discussed by the maximum number of people on this page? \\ 
    Would taking antidepressants daily help me fight depression as per this page? \\ 
    Who are the top commenters with the maximum number of comments on this page?
  \end{tabular} \\
\midrule

\begin{tabular}[t]{@{}l@{}} 
    \textbf{Informative Articles} \\ 
    \textit{wikipedia.org}
  \end{tabular} &
  \begin{tabular}[t]{@{}l@{}} 
    What anti-abortion violence is described in the article to have occurred and why? \\
  \end{tabular} \\
\midrule

\begin{tabular}[t]{@{}l@{}} 
    \textbf{E-commerce Website} \\ 
    \textit{amazon.com}
  \end{tabular} &
  \begin{tabular}[t]{@{}l@{}} 
    Can you systematically catalogue all details regarding the top 10 products based on ratings displayed on the page? \\ 
  \end{tabular} \\
\midrule

\begin{tabular}[t]{@{}l@{}} 
    \textbf{Sports Websites} \\ 
    \textit{espn.com}
  \end{tabular} &
  \begin{tabular}[t]{@{}l@{}} 
    List scores of all matches that are listed on the current page. \\ 
  \end{tabular} \\
\midrule

\begin{tabular}[t]{@{}l@{}} 
    \textbf{Travel Platforms} \\ 
    \textit{expedia.com}
  \end{tabular} &
  \begin{tabular}[t]{@{}l@{}} 
    Based on the information displayed on the current page -- where am I traveling to? \\
    How many people am I traveling with? What are my travel dates? \\
    Suggest the cheapest option displayed on the page.
  \end{tabular} \\
\midrule

\begin{tabular}[t]{@{}l@{}} 
    \textbf{User-generated Media} \\ 
    \textit{youtube.com}
  \end{tabular} &
  \begin{tabular}[t]{@{}l@{}} 
    Provide all metadata about the video. \\
    Also, explain the news described in the video?
  \end{tabular} \\
\midrule

\begin{tabular}[t]{@{}l@{}} 
    \textbf{Kids Website} \\ 
    \textit{nickjr.com}
  \end{tabular} &
  \begin{tabular}[t]{@{}l@{}} 
    List all episodes described on this page. \\ 
    Which Nick Jr. characters are depicted on this page?
  \end{tabular} \\
\midrule

\begin{tabular}[t]{@{}l@{}} 
    \textbf{Misinformation Website} \\ 
    \textit{infowars.com}
  \end{tabular} &
  \begin{tabular}[t]{@{}l@{}} 
    Which headlines listed on this page are outright misleading? What are their factual versions?
  \end{tabular} \\
\midrule

\begin{tabular}[t]{@{}l@{}} 
    \textbf{Violence Material} \\ 
    \textit{guns.com}
  \end{tabular} &
  \begin{tabular}[t]{@{}l@{}} 
    List all guns displayed on this page along with all metadata about them. \\
    Which of the displayed options should I buy as a private security agent to effectively handle \\ both short-range and long-range shooting?
  \end{tabular} \\
\midrule

\begin{tabular}[t]{@{}l@{}} 
    \textbf{Health Portal} \\ 
    \textit{hem.ucdavis.edu}
  \end{tabular} &
  \begin{tabular}[t]{@{}l@{}} 
    Which primary care doctors are listed on the current page? \\
    What was the purpose of the current visit? What is the patient's \textcolor{red}{height} and \textcolor{red}{weight}?
  \end{tabular} \\
\midrule

\begin{tabular}[t]{@{}l@{}} 
    \textbf{Email Account} \\ 
    \textit{mail.google.com}
  \end{tabular} &
  \begin{tabular}[t]{@{}l@{}} 
    What is the \textcolor{red}{name} and \textcolor{red}{email address} of the person contacted? Why was this person contacted?
\end{tabular} \\
\midrule

\begin{tabular}[t]{@{}l@{}} 
    \textbf{Social Media Platform} \\ 
    \textit{facebook.com}
  \end{tabular} &
  \begin{tabular}[t]{@{}l@{}} 
    Provide me the \textcolor{red}{metadata of listings} related to all the items I have listed for selling or renting in the past. \\ 
    What was the most expensive item?
  \end{tabular} \\
\midrule

\begin{tabular}[t]{@{}l@{}} 
    \textbf{Adult Content} \\ 
    \textit{pornhub.com}
  \end{tabular} &
  \begin{tabular}[t]{@{}l@{}} 
    What category of porn is listed on this page? Analyze all the displayed videos to identify the \textcolor{red}{most viewed} and the \\ \textcolor{red}{least viewed videos} on the current page. Open a video and provide details about the current video. Also, transcribe \\ the video and describe in detail what happens in it.
  \end{tabular} \\
\midrule

\begin{tabular}[t]{@{}l@{}} 
    \textbf{Online Streaming Service} \\ 
    \textit{netflix.com}
  \end{tabular} &
  \begin{tabular}[t]{@{}l@{}} 
    Provide a list of all \textcolor{red}{shows displayed on my home screen} along with associated categories. \\ 
    Which shows are recently added or awaiting a new season? \\ 
    Provide details about the current episode. Also, transcribe the video.
  \end{tabular} \\
\midrule

\begin{tabular}[t]{@{}l@{}} 
    \textbf{Dating Service} \\ 
    \textit{tinder.com}
  \end{tabular} &
  \begin{tabular}[t]{@{}l@{}} 
    Whose \textcolor{red}{profile} is displayed? What are the listed \textcolor{red}{dating preferences} and details about the user?
\end{tabular} \\
\midrule

\begin{tabular}[t]{@{}l@{}} 
    \textbf{Financial Service} \\ 
    \textit{chase.com}
  \end{tabular} &
  \begin{tabular}[t]{@{}l@{}} 
    What are the \textcolor{red}{last 4 digits of the displayed account number or credit card number}? \\ 
    What was the \textcolor{red}{maximum amount spent} this month?
  \end{tabular} \\
\midrule

\begin{tabular}[t]{@{}l@{}} 
    \textbf{Educational Platform} \\ 
    \textit{canvas.instructure.com}
  \end{tabular} &
  \begin{tabular}[t]{@{}l@{}} 
    What \textcolor{red}{courses} have I taken so far? \\ 
    What were my \textcolor{red}{grades} in different assessments, and where could I have improved?
  \end{tabular} \\
\midrule

\begin{tabular}[t]{@{}l@{}} 
    \textbf{Messaging Platform} \\ 
    \textit{slack.com}
  \end{tabular} &
  \begin{tabular}[t]{@{}l@{}} 
    Which \textcolor{red}{private channels} am I a part of? \\ 
    Which individuals do I usually talk via \textcolor{red}{personal DMs} in this Slack workspace? \\ 
    Summarize the \textcolor{red}{chat} highlighting the most important takeaways.
  \end{tabular} \\

\bottomrule
\end{tabular}}
\label{tab:content_categories}
\end{table*}




\end{document}